\documentclass{ws-ijqi}
\usepackage[utf8]{inputenc}
\usepackage{thm,amsmath}
\newtheorem{thm}{\bf Theorem}
\newtheorem{prop}[thm]{\bf Proposition}
\newtheorem{rem}[thm]{\bf Remark}
\newtheorem{defin}[thm]{\bf Definition}

\usepackage{mathrsfs} 
\usepackage{amssymb}
\usepackage{mathtools}
\usepackage{braket}
\usepackage{siunitx} 
\usepackage{graphicx}
\usepackage{tensor}
\usepackage{tabularx} 
\usepackage{longtable} 
\usepackage{enumitem}
\usepackage{scalerel,stackengine}
\usepackage{color}
\usepackage{hhline} 
\usepackage{bbold}
\newcommand{\be}{\begin{equation}}
\newcommand{\ee}{\end{equation}}
\newcommand{\bes}{\begin{equation*}}
\newcommand{\ees}{\end{equation*}}
\newcommand{\bs}{\begin{split}}
\newcommand{\es}{\end{split}}
\newcommand{\ba}{\begin{array}}
\newcommand{\ea}{\end{array}}
\newcommand{\eqd}{\vcentcolon=}
\DeclareMathOperator{\tr}{Tr}
\newcommand\mydots{\hbox to 1em{.\hss.\hss.}}
\newcommand\smalldots{\hbox to 0.7em{.\hss.\hss.}}
\def\apeqA{\SavedStyle\sim}
\def\apeq{\setstackgap{L}{\dimexpr.5pt+1.5\LMpt}\ensurestackMath{%
  \ThisStyle{\mathrel{\Centerstack{{\apeqA} {\apeqA}}}}}}
\begin{document}
\title{QUANTUM ENHANCED METROLOGY OF HAMILTONIAN PARAMETERS BEYOND THE CRAM\`ER-RAO BOUND}
\markboth{Luigi Seveso \& Matteo G. A. Paris}{Quantum enahnced metrology
beyond the CR bound}
\author{LUIGI SEVESO, MATTEO G. A. PARIS}
\address{
Dipartimento di Fisica 'Aldo Pontremoli', Universit\`a  di
Milano,
I-20133 Milano, Italy \\
matteo.paris@fisica.unimi.it}
\maketitle
\begin{history}
\received{\today}
\end{history}
\begin{abstract}
This is a tutorial aimed at 
illustrating some recent developments in quantum 
parameter estimation beyond the Cram\`er-Rao bound, as well as 
their applications in quantum metrology. Our starting point is the 
observation that there are situations in classical and 
quantum metrology where the unknown parameter of interest, 
besides determining the state of the probe, is also influencing 
the operation of the measuring devices, e.g. the range of 
possible outcomes.  In those cases, non-regular
statistical models may appear, for which the Cram\`er-Rao theorem does not 
hold. In turn, the achievable precision may exceed the Cram\`er-Rao 
bound, opening new avenues for enhanced metrology. We focus on 
quantum estimation of Hamiltonian parameters and show that an 
achievable bound to precision (beyond the Cram\`er-Rao) 
may be obtained in a closed form for the class of so-called  
controlled energy measurements. Examples of applications of the 
new bound to  various estimation problems in quantum metrology 
are worked out in some details.
\end{abstract}
\section{Introduction}
In the last decade, quantum signals and detectors carved out a place for themselves in mainstream technology. Characterization of those devices at 
the quantum level is thus a crucial ingredient for the development of quantum technologies. Quantum metrology, on the other hand, is the art of estimating the value of one or more parameters of interest, e.g. those characterizing the operation of a device, by exploiting the quantum features of {\em both} the probing system and the measuring apparatus. This second, broadly employed, understanding of the concept has attracted the interest of many researchers, causing a rapid development of the field \cite{q1,q2}. 
\par
Quantum estimation theory (QET) is the  mathematical framework where 
to address optimization of a quantum measurement \cite{q2,q3}. It applies to situations where on is interested in inferring the value of a parameter by performing a 
set of measurements on identical repeated preparations of the system, 
and then processing data in order to estimate the value of the unknown parameter. In turn, the goal of QET is to optimize the overall inference startegy, i.e. the two following steps: 1) the choice of the 
most convenient measurement apparatus and 2) the choice of the most 
convenient estimator, i.e. the data processing able to extract as much information as possible about the parameter of interest. The figure 
of merit used to assess the precision the estimation is the mean 
square error and an inference strategy is deemed optimal if the 
mean square error achieves a minimum. Step number two in the 
above list is classical 
in nature, and amounts to choose a suitable data processing. On the 
other hand, the 
first one is where the quantum nature of physical devices come into play. 
\par
Usually, the choice of the optimal measurement is made by optimizing the figure of merit assuming that the information on the unknown parameter comes from the statistical manifold of possible quantum states of the system only. In other words, one assumes that the measurement apparatus aimed at estimating the parameter does not depend on its value. Such an assumption is necessary to employ standard tools of QET, i.e. the concept of quantum Fisher information and the so-called quantum Cram\`er-Rao theorem.
\par
As a matter of fact, there are relevant estimation problems where the above assumption does not hold. In those cases, an alternative approach is needed  to obtain the ultimate precision bounds, as imposed by quantum mechanics. Relevant examples are provided by statistical models for Hamiltonian parameters, and
by models where the sample space of possible results do depend itself on
the parameter of interest. In order to address those scenarios, novel bounds have been proposed, some of them being tight and achievable. In particular, it has been proved that 
the achievable precision may exceed the Cramer-Rao bound, thus 
opening new avenues for quantum enhanced metrology \cite{seveso_quantum_2017,seveso_estimation_2017}. 
\par
In this tutorial, we review  some recent developments in quantum parameter 
estimation beyond the Cramer-Rao bound, as well as their applications in 
quantum metrology. We focus on 
quantum estimation of Hamiltonian parameters, illustrate the novel 
bound (beyond the Cram\`er-Rao one) for the so-called class of 
controlled energy measurements, and work out in details few examples
of applications, especially those of interest for quantum magnetometry.
In order to place the reader in a position to appreciate the recent developments, we will introduce in details the basic notions of quantum parameter estimation, paying the necessary attention to the mathematical framework where those notions had been developed. In turn, the paper is structured as follows: In Section \ref{s:stat} we provide a brief summary of concepts and notations used in probability theory, whereas Section 
\ref{s:clas} is devoted to classical parameter estimation and Section \ref{qmth} to quantum measurement theory. Quantum parameter estimation is briefly reviewed in Section \ref{s:qett}, whereas non-regular measurements and parameter estimation beyond the quantum Cram\'er-Rao theorem are discussed in Section \ref{pebcr}. Non-regular estimation of general Hamiltonian parameters is the subject of Section \ref{s:nonr}. In particular, we analyze metrological scheme based on controlled energy measurements and present a tight achievable bound for the precision they may achieve. In Sections \ref{algo} and \ref{examples} we discuss metrological applications of the above findings, and work out in details few examples of interest in quantum magnetometry. 
Section \ref{s:out} closes the paper with some concluding remarks.  
\section{Elements of probability theory}\label{s:stat}
The outcome of a random experiment is an \emph{event}. At this stage, an event has no numerical counterpart: it is an abstract subset of a \emph{sample space} $\Omega$. In general, not every possible subset of $\Omega$ constitutes an event. A few desirable requirements are the following: an experiment may have no outcome, so the empty set should be an event; if $A$ is a possible event, then its complement $A^c$, or logical negation, should also be an event; if $A$ and $B$ are events, then their union $A \cup B$, or logical conjunction, should also be an event. Such requirements naturally lead to the introduction of a $\sigma$-algebra structure on the set of events.
\begin{defin}
(\textbf{$\sigma$-algebra}) 
A $\sigma$-algebra $\mathcal A$ on a sample space $\Omega$ is a family of subsets of $\Omega$ having the following properties:
\emph{(\textbf{P1})}: The empty set $\emptyset $ is an element of $ \mathcal A$;\emph{(\textbf{P2})}: If $A$ is an element of $\mathcal A$, then also its relative complement $A^c \in \mathcal A$; \emph{(\textbf{P3})}: If $\{A_i\}_{i=1}^{\infty}$ is a countable collection of elements of $\mathcal A $, then also $\bigcup_{i=1}^\infty A_i \in \mathcal A$.
\end{defin}
The tuple $(\Omega, \mathcal A)$ is called a \emph{measurable space} and the elements of $\mathcal A$ the \emph{measurable sets}. Making use of properties $(\textbf{P2})-(\textbf{P3})$, one may prove that, if $\{A_i\}_{i=1}^{\infty}$ is a countable collection of elements in $ \mathcal A $, then also their countable intersection $\bigcap_{i=1}^\infty A_i \in \mathcal A$. It follows that if $\mathcal A_1$ and $\mathcal A_2$ are two different $\sigma$-algebras on the same sample space
$\Omega$, then their intersection $\mathcal A_1 \cap \mathcal A_2$ is also a $\sigma$-algebra. From this, one may go on to prove that, given any  family of sets $\tau$, there is a unique smallest $\sigma$-algebra containing $\tau$. A case of major interest is when $\tau$ is a topology on $\Omega$, i.e.~the tuple $( \Omega, \tau)$ is a \emph{topological space}.
\begin{defin} (\textbf{topological space}) A topological space $( \Omega, \tau)$ is a set $\Omega$ provided with a topology $\tau$, i.e.~a family of subsets of $\Omega$ having the following properties: \emph{(\textbf{P1})}: Both $\Omega$ and the empty set $\emptyset$ are elements of $\tau$; \emph{(\textbf{P2})}: If $\{T_i\}_{i=1}^{\infty}$ is a countable collection of elements of $\tau $, then also $\bigcup_{i=1}^\infty T_i \in \tau$;
\emph{(\textbf{P3})}: If $\{T_i\}_{i=1}^{n}$ is a finite collection of elements of $\tau$, then also $\bigcap_{i=1}^n T_i \in \tau$.
\end{defin}
The elements of a topology $\tau$ on $\Omega$ are called the \emph{open sets} of $\Omega$. If $\Omega$ is endowed with a topological structure to start with, a $\sigma$-algebra structure can be introduced by taking countable unions, countable intersections and relative complements of its open sets. The resulting $\sigma$-algebra is called the \emph{Borel $\sigma$-algebra} $\mathscr B(\Omega)$: it is the smallest $\sigma$-algebra containing the open sets of $\Omega$.
Once a $\sigma$-algebra structure  $\mathcal A$ has been introduced on $\Omega$, the probability of different events is specified by a \emph{probability measure} $\mu$. The triple $(\Omega, \mathcal A, \mu)$ is called a \emph{probability space}.
\begin{defin}(\textbf{probability space})
A probability space $(\Omega, \mathcal A, \mu)$ is a set $\Omega$ together with a $\sigma$-algebra structure $\mathcal A$ and a probability measure $\mu$, i.e.~a function $\mu: \mathcal A \to [0,1]$ having the following properties:
\emph{(\textbf{P1})}: $\mu(\Omega)=1$; 
\emph{(\textbf{P2})}: If $\{A_i\}_{i=1}^{\infty}$ is a countable collection of mutually disjoint elements of $\mathcal A $, then
\begin{equation}
\mu\left(\bigcup_{i=1}^\infty A_i\right)=\sum_{i=1}^\infty \mu(A_i)\;.
\end{equation}
\end{defin}
With the help of property $(\textbf{P2})$, one may also prove that a probability measure satisfies the following intuitive properties: $\mu(\emptyset)=0$; if $B \subset A$, then $\mu(B)\leq \mu(A)$; for any two events $A,B \in \mathcal A$, $\mu(A\cup B)=\mu(A)+\mu(B)-\mu(A\cap B)$.
Notice that if $(\Omega,\mathcal A)$ is a measurable space and $\mu:\mathcal A\to \overline{\mathbb R}_+$ is a function from the measurable sets to the extended (nonnegative) real line, satisfying property \emph{\textbf{(P2)}}, the triple $(\Omega,\mathcal A,\mu)$ is called a \emph{measure space} and $\mu$ a \emph{measure}. A measure $\mu$ is said to be \emph{finite} if $\mu(\Omega)$ is a finite real
number (it is said \emph{$\sigma$-finite} if $\Omega$ is countable union of measurable sets having finite measure). A probability space is thus equivalent to a measure space with finite measure, normalized according to property \emph{\textbf{(P1)}}. While the random outcomes of an experiment are only required to have a $\sigma$-algebra structure, a \emph{random variable} is needed in order to associate values to elements of $\Omega$.
\begin{defin}\label{defRV}(\textbf{random variable})
Given a probability space $(\Omega, \mathcal A, \mu)$ and a measurable space $(\mathcal X, \mathcal B)$, a random variable is a function $\mathbb X: \Omega\to \mathcal X$ having the following property: if $B\in \mathcal B$, then the preimage of $B$ under $\mathbb X$, i.e.~$\mathbb X^{-1}(B)=\{\omega \in \Omega : \mathbb X (\omega)\in B\}$, is an element of $\mathcal A$.
\end{defin}
Notice that the measurable space $(\mathcal X, \mathcal B)$ in Def.~\ref{defRV} can be naturally made into a probability space, by introducing the probability measure $\nu$ defined via the relation
$\nu(B) = \mu(\mathbb X^{-1}(B))$, where $B$ is any measurable set in $\mathcal B$. In practice, one often blurs the distinction between the two probability spaces $(\Omega, \mathcal A, \mu)$ and $(\mathcal X, \mathcal B, \nu)$, and says that the outcome of a random experiment is a real value $x \in \mathcal X$, rather than an event $A \in \mathcal A$. We will also make use of such abuse of terminology when the distinction can be safely ignored. We add that since,
by definition, a {measurable function} between two measurable spaces is a function such that the preimage of any measurable set is measurable, then 
a random variable can equivalently be defined as a measurable function between  probability spaces.
For most random variables of interest, the image set $\mathcal X$ is a subset of the real line $\mathbb R$. If the subset is finite or countably infinite, the random variable is said to be \emph{discrete}; otherwise, it is a \emph{continuous}. In the following, by a random variable, it will always be meant a \emph{real} random variable, either discrete or continuous. We will also assume that the $\sigma$-algebra $\mathcal B$ is fixed by defining  first a topological structure on $\mathcal X$ (i.e., the subspace topology induced by the real line standard topology) and then a $\sigma$-algebra structure, i.e.~the Borel algebra of $\mathcal X$.
\par
Let us now sketch how to define a notion of integration of a random variable with respect to a probability measure. This is done initially only for \emph{simple} random variables.
\begin{defin} \textbf{(simple random variable)}
A random variable $\mathbb X: \Omega \to \mathcal X$ is simple if $\mathcal X$ is a finite set.
\end{defin}
As a consequence, a simple random variable $\mathbb X$ can be written as $\mathbb X = \sum_{i=1}^n x_i \boldsymbol{1}_{A_i}$, where $\{x_i\}_{i=1}^n$ are real numbers, $\{A_i\}_{i=1}^n$ are elements of $\mathcal A$ and $ \boldsymbol{1}_{A_i}$ is the characteristic function of $A_i$, i.e.
\begin{equation}
\boldsymbol{1}_{A_i}(\omega) \eqd \begin{cases}
1\qquad \text{if $\omega\in A_i$}\\
0\qquad \text{if $\omega\notin A_i$}
\end{cases}\;.
\end{equation}
This representation is, in general, non-unique.
If $\mathbb X$ is a simple random variable, its \emph{expectation} is defined as
\begin{equation}
E(\mathbb X) \eqd \sum_{i=1}^n x_i\, \mu(A_i)\;,
\end{equation}
which can also be denoted by $\int_{\Omega}\mathbb X\, d\mu$. It can be proven that $E(\mathbb X)$ does not depend on the representation.
The next step is to define the expectation of nonnegative random variables. A random variable $\mathbb X$ is nonnegative if it takes only nonnegative values. Two random variables satisfy $\mathbb X \geq \mathbb Y$ if their difference $\mathbb X -\mathbb Y$ is nonnegative. One defines:
\begin{equation}
E(\mathbb X) \eqd  \text{sup}\,(E(\mathbb Y),\; \text{$\mathbb Y$ a simple random variable with $0\leq \mathbb Y \leq \mathbb X$})\;.
\end{equation}
Let us remark that, by definition, $E(\mathbb X)\geq 0$ and that $E(\mathbb X)$ always exists, but might be equal to $+\infty$, even if $\mathbb X$ is everywhere finite.
The final step is to consider an arbitrary random variable $\mathbb X$. Let $\mathbb X^{(+)} = \text{max}(\mathbb X, 0)$ and $\mathbb X^{(-)} = - \text{min}(\mathbb X, 0)$. Thus, $\mathbb X = \mathbb X^{(+)}- \mathbb X^{(-)}$, where $\mathbb X^{(+)}$ and $\mathbb X^{(-)}$ are positive random variables. Then, one defines
\begin{equation}\label{expect}
E(\mathbb X) \eqd E(\mathbb X^{(+)}) - E(\mathbb X^{(-)})\;.
\end{equation}
\par
A random variable $\mathbb X$ is \emph{integrable} if both $E(\mathbb X^{(+)})$ and $E(\mathbb X^{(-)})$ are finite; then, its expectation is given by Eq.~\eqref{expect}. It is easy to check that the set of integrable random variables on a probability space $(\Omega, \mathcal A, \mu)$ is a vector space, denoted by $\mathcal L^1$, with expectation acting as a linear map on it. Notice that, if two random variables satisfy $\mathbb X = \mathbb Y$ almost
surely, i.e.~$\mu(\{\omega \in\Omega:\,\mathbb X(\omega) = \mathbb
Y(\omega)\})=1$, then $E(\mathbb X) = E(\mathbb Y)$.  Therefore, equality almost surely is an equivalence relation, denoted by $\sim$, and equivalent random variables have the same expectation. To remove this redundancy, one introduces the quotient space $L^1\eqd \mathcal
L^1/\!\sim$, whose elements are equivalence classes of almost surely equal random variables. However, by abuse of terminology, one usually still refers to elements of  $L^1$ as random variables. In a similar way, for $1\leq p < \infty$, one defines $\mathcal L^p$ as the vector space of random variables such that $|\mathbb X|^p\in \mathcal L^1$, where $|\mathbb X|\eqd \mathbb X^{(+)}+\mathbb X^{(-)}$. By taking equivalence classes with respect to $\sim$, one then obtains the spaces $L^p$ of $p$-integrable random variables. In the following, we will only need the spaces $L^1$ and $L^2$.
\par
If two random variables are square-integrable, they satisfy the following inequality.
\begin{prop}\label{CSinequality} \textbf{(Cauchy-Schwarz inequality)}
If $\mathbb X, \mathbb Y \in L^2$, then $\mathbb X\cdot \mathbb Y \in L^1$ and
\begin{equation}
|E(\mathbb X \cdot \mathbb Y)| \leq \sqrt{E(\mathbb X^2)E(\mathbb Y^2)}\;.
\end{equation}
\end{prop}
Given square-integrable random variables $\{\mathbb X_i\}_{i=1}^n$ with $\mathbb X_i \in L^2$, one defines their covariance matrix as follows:
\begin{defin} \textbf{(covariance matrix)}
Let $\{\mathbb X_i\}_{i=1}^n$ be a collection of square-integrable random variables in $L^2$. Their covariance matrix is the matrix with entries:
\begin{equation}
\text{\emph{Cov}}(\mathbb X_i,\mathbb X_j) \eqd E\left[(\mathbb X_i - E(\mathbb X_i))(\mathbb X_j - E(\mathbb X_j))\right] \;.
\end{equation}
\end{defin}
In particular, the diagonal elements of a covariance matrix are the variances $\text{\emph{Var}}(\mathbb X_i) \eqd E[(\mathbb X_i - E(\mathbb X_i))^2]$.
As a concluding remark, since the product of two measurable functions is a measurable function and the characteristic function $\boldsymbol{1}_A$ of a set $A$ is measurable if and only if $A$ is measurable, the integral of a random variable on any measurable set $A \in \mathcal A$ is well-defined: one has to take the expectation of the product $\boldsymbol{1}_A \cdot \mathbb X $, i.e. $\int_{A} \mathbb X\, d\mu = \int_{\Omega} \boldsymbol{1}_A \cdot \mathbb X\, d\mu$.
\par
We now introduce the concept of \emph{probability density} of a random variable. As discussed before, a random variable $\mathbb X$ on a probability space $(\Omega, \mathcal A, \mu)$ gives rise to a probability space $(\mathcal X, \mathcal B, \nu)$, where $\mathcal X \subseteq \mathbb R$, $\mathcal B$ is the Borel algebra generated by the natural topology of $\mathcal X$ and $\nu$ is a probability measure. Notice that there are already two natural notions of a measure on $\mathcal X$: the \emph{Lebesgue measure} (if $\mathcal X$ is an uncountable
subset of $\mathbb R$) and the \emph{counting measure} (if $\mathcal X$ is a countable subset). The measure $\nu$ can always be expressed in terms of either the Lebesgue measure or the counting measure, provided it satisfies a technical assumption, which is contained in the following definition.
\begin{defin}\textbf{(absolutely continuous measures)}
If $\nu$ and $\nu'$ are any two measures with the same $\sigma$-algebra $\mathcal B$ of subsets of $\mathcal X$, then $\nu$ is said to be absolutely continuous with respect to $\nu'$, denoted $\nu \ll \nu'$, if $\nu(B)=0$ for any $B\in \mathcal B$ such that $\nu'(B)=0$.
\end{defin}
We henceforth assume that, if $\mathbb X$ is a continuous random variable, $\nu$ is absolutely continuous with respect to the Lebesgue measure, i.e.~it agrees with the Lebesgue measure on any set with Lebesgue measure zero. If instead $\mathbb X$ is discrete, every probability measure $\nu$ is already absolutely continuous with respect to the counting measure (since the counting measure vanishes only on the empty set and $\nu(\emptyset)=0$ always). The following theorem applies to any two absolutely continuous measures.
\begin{thm}\textbf{(Radon-Nikodym)}
Let $\nu$ and $\nu'$ be two $\sigma$-finite measures on the same measurable space $(\mathcal X, \mathcal B)$ such that $\nu \ll \nu'$. Then:
\emph{\textbf{(T1)}}: There exists a measurable function $h:\mathcal X \to \mathbb R_+$ such that, for all $B\in \mathcal B$,
\begin{equation}\label{RNproperty}
\nu(B) = \int_B h \,d\nu'\;.
\end{equation}
\emph{\textbf{(T2)}}: Such a function $h$ is almost unique: any two functions satisfying Eq.~\eqref{RNproperty} can differ only on sets of measure zero with respect to $\nu'$. \emph{\textbf{(T3)}}: $h$ is integrable with respect to $\nu'$ if and only  $\nu$ is a finite measure.
\end{thm}
The function $h$ is called the Radon-Nikodym derivative of $\nu$ with respect to $\nu'$, denoted $h=d\nu/d\nu'$. It allows to convert between the two measures by means of the symbolic identity $d\nu = h\, d\nu'$. Given the above
let us introduce the following definition
\begin{defin}
Let $\mathbb X$ be a random variable with probability space $(\mathcal X, \mathcal B, \nu)$. \emph{\textbf{(D1)}}: If $\mathbb X$ is continuous, its probability density function (p.d.f.) is the Radon-Nikodym derivative of $\nu$ with respect to the Lebesgue measure, i.e.~$p=d\nu/dx$. \emph{\textbf{(D2)}}: If $\mathbb X$ is discrete, its probability mass function (p.m.f.) is the Radon-Nikodym derivative of $\nu$ with respect to the counting measure, i.e.~$p=d\nu/d\#$.
\end{defin}
Knowledge of the p.d.f.~$p$ (\emph{resp}., the p.m.f.) fully characterizes a random experiment whose outcomes are described by a random variable $\mathbb X$, since the probability of any event $B$ can be obtained by integrating $p$ on $B$ with respect to the Lebesgue measure (\emph{resp}., the counting measure) by means of Eq.~\eqref{RNproperty}.
\section{Classical parameter estimation}\label{s:clas}
Let us consider a random experiment, whose outcomes are described 
by a random variable $\mathbb X$, with probability space $(\mathcal X, \mathcal B, \nu)$ and probability density $p$. The task is to reconstruct $p$, which is referred to as the \emph{true} probability density, starting from $N$ independent sample points or observations of $\mathbb X$ (in the following, a sample point is denoted by a lowercase letter, e.g.~$x\in \mathcal X$, whereas a sample of $N$ observations  by a boldface letter, e.g.~$\boldsymbol x \in \mathcal X^{\times N}$).
There are many ways to approach the problem of learning $p$ but, if the functional form of $p$ is already known, or can be guessed with reasonable accuracy, a \emph{parametric approach} is quite natural. The true probability density $p$ is assumed to belong to a parametric family of
probability densities $\{p_\theta\}_{\theta\in\Theta}$, where $\Theta \subset \mathbb R^m$ is the \emph{parameter space}. It is also assumed that there exists a suitable choice $\theta^* \in \Theta$ such that $p_{\theta^*}=p$. In this way, all lack of knowledge about $p$ is reduced to lack of knowledge about the \emph{true}  parameter $\theta^*$ -- a considerable simplification of the problem.
\begin{defin}\textbf{(classical statistical model)}
A classical statistical model $S$ is a family of probability densities on $\mathcal X$ parametrized by $m$ real parameters $ \theta \in  \Theta \subset \mathbb R^m$:
\begin{equation}
S = \{ p_\theta \,:\,\theta = (\theta^1,\theta^2,\dots, \theta^m) \in \Theta\}\;,
\end{equation}
where the parametrization map $\theta \to p_\theta$ is injective, the support $\mathcal X$ is parameter-independent and $p_\theta$ can be differentiated as many times as needed with respect to the parameters, i.e.~all possible derivatives $\partial_1^{k_1} \dots \partial_m^{k_m} p_\theta$ (where $\partial_i$ is short for $\partial_{\theta^i}$) exist.
\end{defin}\noindent
Notice that if $\mathcal X$ is countable, then $p_\theta$ is a p.m.f. normalized such that
\begin{equation}
\sum_{x\in\mathcal X} p_{x,\theta} = 1\;, \qquad \forall \theta \in \Theta\;.
\end{equation}
If $\mathcal X$ is uncountable, then $p_\theta$ is a p.d.f. normalized such that
\begin{equation}
\int_{\mathcal X} p_{\theta}(x)\,dx = 1\;,  \qquad \forall \theta \in \Theta\;.
\end{equation}
In the following, we will employ the notation for continuous variables; for discrete variables, one should replace the Lebesgue measure $dx$ by the counting measure $d\#$.
\par
Given a statistical model $S=\{p_\theta\}_{\theta\in\Theta}$, the map $\varphi: S \to \mathbb R^m$ defined by $\varphi(p_\theta)=\theta$ can be considered as providing a coordinate system for $S$. If $\psi$ is a smooth reparametrization which maps $\Theta\to \Theta'$, nothing prevents using $\psi(\theta)=\theta'$ as the new parameters, so that the model is rewritten as
$S=\{p_{\psi^{-1}(\theta')}\,:\,\theta' \in \Theta'\}$. This defines the structure of a differentiable manifold on $S$, with different parametrizations representing different coordinate systems. Moreover, a Riemannian metric can be defined on the statistical manifold $S$ as follows.
\begin{defin} \textbf{(Fisher information)}\label{fidefin}
Let $S$ be a statistical model. Given a point $\theta$, the (classical) Fisher information matrix $\mathcal F_C(\theta)$ at that point is the matrix having $(i,j)^{\text{th}}$ element
\begin{equation}\label{FIDefin}
  [\mathcal F_C(\theta)]_{ij} = \int_{\mathcal X} dx\,p_{\theta}(x)\, \partial_i \log p_{\theta}(x)\;\partial_j \log p_{\theta}(x)  \;.
\end{equation}
\end{defin}\noindent
When $m=1$ and only one parameter $\theta=\theta^1$ is present, $\mathcal F_C(\theta)$ is referred to as the Fisher information (FI). For $m>1$, $\mathcal F_C(\theta)$ is indeed a symmetric $m\times m$ real matrix. It is always positive semi-definite and, in particular, positive-definite if and only if for every $\theta\in \Theta$ the elements of the set $\{\partial_1 p_\theta,\dots,\partial_m p_\theta\}$ are linearly independent. Moreover, $\mathcal F_C(\theta)$ has the correct transformation properties of a $(0,2)$
tensor under reparametrizations \cite{amari_methods_2007}.  
It follows that $\mathcal{F}_C (\theta)$ provides a Riemannian metric on $S$.
There is a precise sense in which the Fisher geometry, i.e.~the geometry implied by the Fisher information metric, is the only possible geometry on a statistical manifold. To explain this, we introduce the notion of a \emph{statistic}.
\begin{defin} \textbf{(statistic)}
Given a random variable $\mathbb X$ and a function $T:\mathcal X \to \mathcal Y$ which maps $x \to y = T(x)$, a statistic based on $T$ is the random variable $\mathbb Y = T(\mathbb X)$.
\end{defin}\noindent
If $\mathbb X$ is associated with a statistical model $S=\{p_\theta\}_{\theta \in\Theta}$, then a statistic $T$ gives rise to a model $S_T=\{q_\theta\}_{\theta\in\Theta}$ associated with $\mathbb Y = T(\mathbb X)$. A statistic is said to be \emph{sufficient} if the two models are related as follows: $p_\theta(x) = h(x)\, q_\theta(y(x))$, $\forall x \in \mathcal X$, i.e.~all dependence on the parameter $\theta$ is contained in
$q_\theta$. Intuitively, a sufficient statistic leads to no loss of information about $\theta$. Notice that a one-to-one function is always a sufficient statistic, but  there exist sufficient statistics which are not one-to-one functions. We now have the following theorem.
\begin{thm} \textbf{(pre Cram\`er-Rao)} The Fisher information matrix $\mathcal F_C^{(T)}$ of the statistical model $S_T$ induced by a statistic $T$ satisfies the monotonicity property $\mathcal F_C^{(T)} \leq \mathcal F_C$ (where $\mathcal F_C$ is the Fisher information matrix of the original model $S$). The previous inequality must be interpreted in the sense that the difference $\mathcal F_C^{(T)}-\mathcal F_C$ is a positive semi-definite matrix. Equality holds if and only if $T$ is a sufficient statistic.
\end{thm}\noindent
A Riemannian metric satisfying the monotonicity property is said to be a monotone metric. Monotone metrics are the natural metrics on classical statistical models: they reflect the fact that the points of the manifold are probability distributions and distances between points can only contract under any information processing. In this regard, the following theorem \cite{campbell_extended_1986,cencov_algebraic_1978,ay_information_2015} singles out the Fisher information metric as the only natural metric on statistical manifolds.
\begin{thm} \textbf{(Chentsov)}
The Fisher information metric $\mathcal F_C$ is the essentially unique monotone Riemannian metric on a classical statistical model, in the sense that any other such metric is a scalar multiple of $\mathcal F_C$.
\end{thm}\noindent
Chentsov's theorem establishes a first link between the statistical properties of parametric models and the geometry defined by the Fisher metric. A further link comes from the (classical) Cram\'er-Rao theorem, which we now introduce.
\par
Let us now return to the problem of estimating the true parameter $\theta^*$ from a sample $\boldsymbol x \in \mathcal X^{\times N}$. To this end, we introduce the following definition.
\begin{defin}\textbf{(estimator)}
An estimator $\hat \theta^{(N)}: \mathcal X^{\times N}\to \Theta$ is a random variable from the sample space $\mathcal X^{\times N}$ to the parameter space $\Theta$. In particular, \emph{\textbf{(D1)}}: An unbiased estimator is an estimator satisfying $E_\theta(\hat \theta^{(N)})=\theta$, $\forall \theta \in \Theta$, where $E_\theta(\cdot)$ denotes expectation with respect to $p_\theta$, i.e.
\begin{equation}
E_\theta(\hat \theta^{(N)}) = \int_{\mathcal
X^{\times N}} dx_1\dots dx_N\, p_\theta(x_1)\dots p_\theta(x_N)\,
\hat \theta^{(N)}(\boldsymbol x)\;.
\end{equation}
\emph{\textbf{(D2)}}: A locally unbiased estimator is an estimator which is unbiased at $\theta=\theta^*$, i.e.
\begin{equation}
  E_{\theta^*}(\hat \theta^{(N)})=\theta^*\;,
\end{equation}
and, moreover, satisfies
\begin{equation}
  \partial_i E_\theta[(\hat \theta^{(N)})^j]\big|_{\theta=\theta^*}=\delta_i^j\;.
\end{equation}
\emph{\textbf{(D3)}}: An asymptotically unbiased estimator is an estimator such that
\begin{equation}
  \lim_{N\to\infty} E_\theta(\hat \theta^{(N)})=\theta\;.
\end{equation}
\end{defin}
A typical (classical) estimation protocol consists in sampling data 
$\boldsymbol x \in \mathcal X^{\times N}$ and the processing them using 
an estimator $\hat \theta^{(N)}$, finally providing an estimate  $\hat \theta^{(N)}(\boldsymbol x)$ of the true value. If the estimator is unbiased, the estimate will fluctuate around the true value $\theta^*$ over many independent repetitions of the protocol. To quantify the performance of an estimator,  it is usual to take as a figure of merit its mean square error:
\begin{equation}
[\text{MSE}(\hat \theta^{(N)})]_{ij}\eqd E_\theta([(\hat \theta^{(N)})^i-\theta][(\hat \theta^{(N)})^j-\theta])\;.
\end{equation}
Estimators with a smaller MSE are said to perform better than estimators with a larger one. Notice that for unbiased estimators, the MSE matrix coincides with the covariance matrix $[\text{Cov}(\hat \theta^{(N)})]_{ij}$. The following theorem provides a lower bound to the covariance matrix of unbiased estimators \cite{rao_information_1992,cramer_mathematical_2016}.
\begin{thm}\label{CRthm}\textbf{(Cram\'er-Rao)}
If $S$ is a classical statistical model and $\hat \theta^{(N)}$ an unbiased estimator, its covariance matrix is bounded from below as follows:
\begin{equation}\label{ineqCR}
  \text{\emph{Cov}}(\hat \theta^{(N)}) \geq  \frac{1}{N}\,[\mathcal F_C(\theta)]^{-1}\,,
\end{equation}
where $\mathcal F_C$ is the Fisher information matrix of $S$.
\end{thm}
The proof of Thm.~\ref{CRthm} amounts to an application of the Cauchy-Schwarz inequality of Prop.~\ref{CSinequality}.
Under the weaker assumption that $\hat \theta^{(N)}$ is only locally unbiased, inequality \eqref{ineqCR} still holds, but only at $\theta = \theta^*$.
Notice that the Cram\'er-Rao theorem only provides a lower-bound: it does not guarantee that an estimator achieving the bound actually exists. If such an estimator exists, it is said to be \emph{efficient}. An efficient estimator is the best unbiased  estimator, since it minimizes the MSE among all unbiased estimators. Unfortunately, efficient estimators exist only under special circumstances (when the statistical model is of the exponential type and the parameters are its natural parameters, see e.g.~Ref.~\cite{casella_statistical_2002}). Finding the best unbiased estimator becomes then a non-trivial task.
The situation improves in the asymptotic limit of a large number of samples. Let us remark that unbiasedness is a strong condition: for some models there exists no such estimator. A far more reasonable condition is that of consistency. A \emph{consistent} estimator is such that, in the limit $N\to \infty$, its probability density becomes concentrated around $\theta$, i.e. $\forall \epsilon > 0$ and $\forall \theta \in\Theta$, $\lim_{N\to \infty} \text{Pr}_\theta (|\hat \theta^{(N)}-\theta|>\epsilon)=0$, where $\text{Pr}_\theta(\cdot)$ denotes the probability of an event
computed with respect to $p_\theta$. Under mild conditions (e.g. that $\text{Cov}(\hat\theta^{(N)})$ is uniformly bounded with respect to the number of samples $N$), one can prove that a consistent estimator is asymptotically unbiased, i.e.~$\lim_{N\to \infty} E_\theta(\hat \theta^{(N)})=\theta$, and satisfies $\lim_{N\to \infty}\partial_i E_\theta[(\hat \theta^{(N)})^j]=\delta^j_i$. With the help of the last two properties, one can prove the following asymptotic version of the Cram\'er-Rao theorem:
\begin{equation}
  \lim_{N\to \infty}N \cdot \text{Cov}(\hat \theta^{(N)}) \geq [\mathcal F_C(\theta)]^{-1}\;.
\end{equation}
A consistent estimator achieving equality is said to be \emph{asymptotically efficient}. Remarkably, asymptotically efficient estimators always exist, e.g. the maximum-likelihood estimator and Bayes estimators are asymptotically efficient \cite{casella_statistical_2002}. In conclusion, at the classical level and in the asymptotic regime $N\gg 1$, the optimal protocol consists in collecting a sample and processing it via an asymptotically efficient estimator; the asymptotic optimal rate at which distinct values of the parameters can be distinguished is given by the inverse Fisher information. 
\section{Quantum measurement theory}\label{qmth}
The outcomes of a quantum experiment are  probabilistic. This means that there must exist a suitable probability measure $\nu_\rho^{(\mathcal M)}$ such that, if $(\mathcal X, \mathcal B)$  is the measurable space of outcomes (where $\mathcal X \subseteq \mathbb R$ is the sample space and $\mathcal B$ the $\sigma$-algebra induced by the natural topology of $\mathcal X$), then the probability of any event $B\in \mathcal B$ is $\nu_\rho^{(\mathcal M)}(B)$. The main difference compared with the classical case is that
$\nu_\rho^{(\mathcal M)}$ is not arbitrary, but is a specific function of both the state of the system $\rho$ and the measurement $\mathcal M$. The mapping
$(\rho,\mathcal M)\to \nu_\rho^{(\mathcal M)}$ is given by Born's rule.
We will deal exclusively with \emph{finite-dimensional} quantum systems, with Hilbert space $\mathcal H =\mathbb C^d$. A state is a \emph{density matrix} $\rho\in \mathsf{Her}_d^+(\mathbb C)$, i.e.~an Hermitian positive semi-definite matrix, usually normalized such that $\tr(\rho)=1$. The set $\mathcal S(\mathcal H)$ of all possible density operators on $\mathcal H$ is a convex set. Its extremal elements are the pure states $\ket{\psi} \bra{\psi}$, with
$\ket{\psi}\in \mathcal H$ such that $\braket{\psi|\psi}=1$.
 The \emph{Hamiltonian} matrix $H\in\mathsf{Her}_d(\mathbb C)$ completely determines the dynamics of the system (assuming it is isolated from any external environment). That is, if $U_t \eqd \text{exp}(-itH)$ is the matrix exponential of $H$ and $\rho_0$ is the state at time $t=0$,  then the state of the system at any subsequent time $t$ is $\rho_t\eqd U_t \rho_0 U_t^\dagger$.
A measurement on a quantum system can be described at three different levels of details. We begin with the first level, which is the more coarse-grained of the three.
 \begin{itemize}
 \item[\textbf{(L1)}] \textbf{POVM description}: At this level, a measurement $\mathcal M$ is a mapping that associates to any event $B \in \mathcal B$ a positive semi-definite  operator $\mathcal M(B)\in \mathsf{Her}^+_d(\mathbb C)$. A few
 natural requirements are that  $\mathcal M(\emptyset) = \mathbb{0}_d$; $\mathcal M(\mathcal X)=\mathbb{1}_d$; if $\{B_i\}_{i=1}^n$ are mutually disjoint measurable sets such that $\bigcup_{i=1}^n B_i\eqd B\in \mathcal B$, then $\mathcal M(B)
 = \sum_{i=1}^\infty \mathcal M(B_i)$. These properties imply that $\mathcal M$ is a positive-operator valued (probability) measure (POVM) on $(\mathcal X,\mathcal B)$. In particular, they imply that if $\{B_i\}_{i=1}^n$ are mutually disjoint and $\bigcup_{i=1}^n B_i = \mathcal X$, then $\sum_{i=1}^n \mathcal M(B_i)=\mathbb I_d$. Apart for this normalization condition and for being non-negative, the operators $\mathcal M( B)$ are completely arbitrary.

The link between a measurement $\mathcal M$ and the probability measure $\nu_\rho^{(\mathcal M)}$ is provided by Born's rule, i.e.
\begin{equation}\label{BRule}
\nu_\rho^{(\mathcal M)} (B) = \tr(\rho\, \mathcal M(B))\;.
\end{equation}
It can be proven that Born's rule is actually the unique possibility under a few reasonable assumptions \cite{gleason_measures_1957}. Eq.~\eqref{BRule} completely determines the statistics of any quantum experiment.

If $\mathcal X$ is a countable sample space, one defines  the \emph{probability operators} $\{\Pi_x\}_{x\in\mathcal X}$ of a given measurement as follows: $\Pi_x \eqd \mathcal M(x)$. The probability operators are sufficient to compute the probability of any other event. A special case is when each $\Pi_x$ is a projector $P_x$, i.e.~$P_x^2=P_x$. One can then associate to the measurement an Hermitian operator $X = \sum_{x\in\mathcal X}x P_x$, also called an \emph{observable}. Vice versa, every Hermitian operator gives rise to a projective measurement via its eigendecomposition. An example is the Hamiltonian: a projective measurement over its eigenstates
$\{\ket{\xi_j}\}_{j=0}^{d-1}$ is called an energy measurement.
\item[]
\item[\textbf{(L2)}] \textbf{Instrument description}: A POVM description assigns probabilities to measurement outcomes, but  does not specify how the state of the system is  modified as a result of the measurement. However, quantum measurements can have dynamical effects: if the measurement is non-destructive, the state of the system is updated depending on the outcome. This requires introducing an \emph{instrument}.
Formally, an instrument $\mathcal I$ is a mapping $\mathcal B \to \mathcal T(\mathcal H)$, where $\mathcal T(\mathcal H)$ denotes the set of bona fide quantum operations on the system (i.e.~completely-positive, trace preserving maps). If $B\in\mathcal B$ is the observed event, then the state of the system after the measurement is, by definition, $\mathcal I_B(\rho)$. Assuming $\mathcal X$ is countable, it is enough to consider the set $\{\mathcal I_x\}_{x\in\mathcal X}$. It can be proven \cite{kraus_states_1983} that the most general form for $\mathcal I_x$ is as follows,
\begin{equation}
  \mathcal I_x(\rho) = \frac{\sum_{j=1}^n M_x^{(j)} \rho\, {M_x^{(j)}}^\dagger}{\text{Pr}(x)}\;,
\end{equation}
where the operators $M^{(j)}_x$ are called \emph{measurement operators} and $\text{Pr}(x)\eqd \tr(\rho\, \Pi_x)$. Since the post-measurement state $\mathcal I_x(\rho)$ must be normalized, one has the identification
\begin{equation}
  \Pi_x = \sum_{j=1}^n {M^{(j)}_x}^\dagger M^{(j)}_x\;.
\end{equation}
In particular, if $n=1$, $\forall x\in\mathcal X$, the measurement is said to be \emph{fine-grained}. Notice that, in general, many different instruments correspond to the same positive-operator valued measure. This is true even for fine-grained measurements, since the condition $\Pi_x = M_x^\dagger M_x$ is solved by $M_x= U_x \sqrt{\Pi_x}$, where $\sqrt{\Pi_x}$ is the principal square-root of $\Pi_x$ but $U_x$ is an arbitrary unitary operator. If the measurement is fine-grained and $U_x=\mathbb I_d$, $\forall x\in\mathcal X$, the measurement is said to be \emph{bare} and the corresponding instrument is known as the \emph{L\"uders instrument}.
\item[]
\item[\textbf{(L3)}] \textbf{Measurement model description}: This is the most detailed level of description of a measurement and is obtained by explicitly modelling the interaction between the system and the measuring apparatus. It is assumed that the system is coupled to an ancillary system with Hilbert space $\mathcal H_A$; the ancilla is prepared in an initial state $\eta \in\mathcal S(\mathcal H_A)$; the two systems evolve together for an interaction time $t_{int}$ via a quantum channel $\mathcal E^{(t_{int})}\in \mathcal T(\mathcal H \otimes \mathcal H_A)$; finally, an observable $X=\sum_{x\in\mathcal X}x P_x$ on $\mathcal H_A$ is
measured, producing an outcome $x\in\mathcal X$. A \emph{measurement
model} is therefore a quadruple $(\mathcal H_A,\eta,\mathcal E^{(t_{int})},X)$. It gives rise to a positive-operator valued measure via the relation:
\begin{equation}\label{cond1}
  \tr(\rho\, \Pi_x) = \tr[\mathcal E^{(t_{int})}(\rho\otimes \eta)\,\mathbb I_d \otimes P_x]\;.
\end{equation}
Moreover, it defines an instrument via
\begin{equation}\label{cond2}
  \mathcal I_x(\rho)= \frac{\tr_A[\mathcal E^{(t_{int})}(\rho\otimes \eta)\,\mathbb I_d \otimes P_x]}{\text{Pr}(x)}\;,
\end{equation}
where $\tr_A(\cdot)$ denotes the partial trace over the ancilla's degrees of freedom.
Clearly, many measurement models can lead to the same instrument. In fact, Ozawa's theorem \cite{ozawa_quantum_1984} states that  one can recover all possible instruments just by considering measurement models $(\mathcal H_A,\eta,\mathcal E^{(t_{int})},X)$ where $\eta$ is pure, $\mathcal E^{(t_{int})}$ is a unitary channel and each $P_x$ is rank-1. More precisely, let $H_A$ be the free Hamiltonian of the ancillary system and $H_{I}$ the interaction Hamiltonian between the system and the apparatus. Let $\eta = \ket{\phi}\bra{\phi}$ be the initial preparation of the ancilla. Then, the unitary channel $\mathcal U^{(t)}$ generated by the total Hamiltonian $H_{T} = H + H_A+ H_{I}$ acts  as follows:
\begin{equation}
  \rho\otimes \ket{\phi}\bra{\phi} \to \mathcal U^{(t)}(\rho\otimes \ket{\phi}\bra{\phi})\eqd U_t\, \rho\otimes \ket{\phi}\bra{\phi} U_t^\dagger\;,\qquad U_t\eqd e^{-it H_T}\;.
\end{equation}
From conditions \eqref{cond1} and \eqref{cond2}, one may prove that the measurement operators $M_x$  and probability operators $\Pi_x$ take the following form, respectively,
\begin{equation}\label{mmPOVM}
  M_x = \braket{x|U_{t_{int}}|\phi}\;,\qquad\quad  \Pi_x = \braket{\phi|U_{t_{int}}^\dagger\;\mathbb I_d \otimes P_x\, U_{t_{int}}|\phi}\;.
\end{equation}
\end{itemize}
\section{Quantum parameter estimation}\label{s:qett}
By analogy with the classical case, a \emph{quantum statistical model} is defined as follows \cite{qp1,qp2}.
\begin{defin}\textbf{(quantum statistical model)}
Given a quantum system described in the Hilbert space $\mathcal H$, 
a quantum statistical model $S$ is a family of states, i.e. density operators, in $\mathcal S(\mathcal H)$ labeled by $m$ real parameters $\theta \in \Theta \subset \mathbb R^m$:
\begin{equation}
  S=\{\rho_\theta\,:\,\theta=(\theta^1,\theta^2,\dots,\theta^m)\in\Theta\}\;,
\end{equation}
where the parametrization map $\theta\to \rho_\theta$ is injective, the rank $\text{\emph{rk}}(\rho_\theta)$ is parameter-independent and $\rho_\theta$ can be differentiated as many times as needed with respect to the parameters.
\end{defin}\noindent
A quantum statistical model typically arises in this way: the system is prepared at time $t=0$ in an initial state $\rho_0$ and then goes through a quantum channel $\mathcal E_{\theta^*}\in\mathcal T(\mathcal H)$, which depends on the true value $\theta^*$ of one or more parameters. The associated model is defined as $\rho_\theta \eqd \mathcal E_\theta(\rho_0)$, with $\theta\in\Theta$  and $\Theta$ containing, by assumption, the true value $\theta^*$.
The mapping $\rho_0 \to \mathcal E_\theta(\rho_0)$ is called the \emph{dynamical encoding}. A typical example is the unitary channel generated by the system's Hamiltonian, i.e.
\begin{equation}
  \rho_\theta = U_{t} \rho_0 U_{t}^\dagger\;,\qquad U_{t}= e^{-it H_\theta}\;.
\end{equation}
 The parameter $\theta$ is usually referred to as a \emph{Hamiltonian parameter}. One then further distinguishes between Hamiltonian 
 \emph{shift} or \emph{phase} parameters and \emph{general} parameters. 
In the first case, the parameter is just and overall multiplicative 
constant, i.e.~$H_\theta = \theta\, G$, i.e. it appears linearly in the Hamiltonian. In the second case, the parameter may appear in any way, e.g. 
non-linearly, and the eigenvectors $\ket{\xi_{j,\theta}}$ of $H_\theta$ generally depend in general on $\theta$.
Dynamical encoding is not, however, the only possibility. For certain models, the encoding is static. A typical example is that of a \emph{thermal  model}, describing the equilibrium state of a quantum system in contact with a thermal bath,
\begin{equation}\label{tbath}
  \rho_\beta = \frac{e^{-\beta H}}{\tr(e^{-\beta H})}\;,
\end{equation}
where the parameter, conventionally denoted by $\beta$, is the inverse temperature of the bath and $H$ is the Hamiltonian of the system.
In both cases, given a quantum statistical model $S=\{\rho_\theta\}_{\theta\in\Theta}$, performing a measurement with probability operators $\{\Pi_x\}_{x\in\mathcal X}$ gives rise to a classical statistical model, via the relation $p_{\theta}(x) = \text{Pr}_\theta(x) = \tr(\rho_\theta \Pi_x)$
(where the sample space $\mathcal X$ is henceforth assumed to be countable). Notice that the choice of the measurement to perform is  an additional degree of freedom the experimentalist is called to optimize upon, which is not present in the classical case. Furthermore, if the encoding is dynamical, one also has to optimize over the initial state of the probe $\rho_0$. As a consequence, the search for optimal quantum estimation protocols is considerably more complicated.
\par
A quantum statistical model can be naturally given the structure of a differentiable manifold. Whereas in the classical case there is a fundamentally unique metric, in the quantum case non-commutativity breaks uniqueness and, in fact, leads to an infinite number of possible metrics. Notice that monotonicity now translates into the requirement that, for any completely-positive, trace-preserving map $\mathcal E\in\mathcal T(\mathcal H)$, the difference between the metric on the original statistical model $\{\rho_\theta\}_{\theta\in\Theta}$ and on the derived model $\{\mathcal E(\rho_\theta)\}_{\theta\in\Theta}$ is positive semi-definite. In the quantum case, all possible monotone Riemannian metrics have been classified by Petz \cite{petz_1996_monotone}. Each such metric is in one-to-one correspondence with an operator monotone function, which in turn is one-to-one related to an operator mean. We give the following definition:
\begin{defin}\textbf{(operator mean)}
  An operator mean $\mathfrak{m}:  \mathsf{Her}^+_d \times \mathsf{Her}^+_d \to \mathsf{Her}^+_d$ is a function such that, for any positive semi-definite operators $A,B,C,D$:
\textbf{\emph{(P1)}}: $\mathfrak{m}(A,A)=A$; 
\textbf{\emph{(P2)}}: $\mathfrak{m}(\alpha A,\alpha A) = \alpha A$, $\forall \alpha \in \mathbb R$; 
\textbf{\emph{(P3)}}: $A\geq C,\,B\geq D\implies \mathfrak{m}(A,B)\geq \mathfrak{m}(C,D)$; 
\textbf{\emph{(P4)}}: $\mathfrak{m}(UAU^\dagger,UBU^\dagger)= U\, \mathfrak{m}(A,B) \,U^\dagger$, $\forall\, U$ unitary; 
\textbf{\emph{(P5)}}: $\mathfrak{m}(A,B)=\mathfrak{m}(B,A)$.
\end{defin}
Any function aspiring to be a mean for positive semi-definite matrices should intuitively satisfy conditions \textbf{(P1)} through \textbf{(P5)}. The following proposition fully characterizes the family of operator means. 
\begin{prop}
Every operator mean can be written in the form
\begin{equation}
  \mathfrak m^{(f)}(A,B) = \sqrt A\,f\left(\frac{1}{\sqrt A} B \frac{1}{\sqrt A}\right)\,\sqrt A\;,
\end{equation}
where $f$ is an operator monotone function (i.e.~a function such that, $\forall A,\,B\in\mathsf{Her}^+_d$, $A\geq B\implies f(A)\geq f(B)$) with the constraints $f(1)=1$ and $f(1/x)=f(x)/x$. Vice versa, any such function gives rise to an operator mean.
\end{prop}
Each quantum monotone metric is now put in one-to-one correspondence with a suitable operator mean via Petz's classification theorem.
\begin{thm} \textbf{(Petz \cite{petz_1996_monotone})}
  If $S=\{\rho_\theta\}_{\theta \in \Theta}$ is a quantum statistical model such that, $\forall \theta \in\Theta$, $\rho_\theta$ is full-rank, the generic monotone Riemannian metric on $S$ is of the form:
  \begin{equation}\label{petzclass}
    [\mathcal F_Q^{(f)}(\theta)]_{ij} = \tr(\partial_i \rho_\theta \,\mathcal J^{-1}\partial_j \rho_\theta)\;,
  \end{equation}
  where $\mathcal  J$ is the superoperator $\mathcal  J = R\, f(L R^{-1})$, $f$ is an operator-monotone function satisfying $f(1) =1$ and $f(1/x) = f(x)/x$, and $L$ (\emph{resp.} $R$) is the left (\emph{resp.} right) multiplication superoperator, which by definition acts on $\eta\in\mathcal S(\mathcal H)$ as follows:
  \begin{equation}
   L(\eta) =  \rho_\theta \eta\;,\qquad R(\eta) =  \eta \rho_\theta\;.
  \end{equation}
\end{thm}\noindent
One may rewrite \eqref{petzclass} more expressively by introducing the \emph{logarithmic derivative} operators $L^{(f)}_{i,\theta}$ which  satisfy the following relations:
\begin{equation}\label{LDDef}
 \partial_i \rho_\theta = \mathcal J L^{(f)}_{i,\theta}\;,\qquad i\in\{1,\dots,m\}\;.
\end{equation}
The metric $\mathcal F_Q^{(f)}$ can therefore be rewritten as
\begin{equation}\label{qMetric}
[\mathcal F_Q^{(f)}(\theta)]_{ij} = \tr[\partial_i \rho_\theta L^{(f)}_{j,\theta}] = \tr[\mathcal J(L^{(f)}_{i,\theta}) L^{(f)}_{j,\theta}]\;.
\end{equation}
For each choice of an operator monotone function $f$, one obtains a corresponding monotone metric.
\begin{itemize}
\item[\textbf{(M1)}] Let us consider the operator monotone function $f_{\text{ari}}(x) = (1+x)/2$. The corresponding operator mean is the \emph{arithmetic mean} since, if $A,B$ are commuting matrices, then $\mathfrak m^{(f_{\text{ari}})}=(A+B)/2$. The logarithimic derivative operator $L_{i,\theta}^{(f_{\text{ari}})}$ satisfies, from Eq.~\eqref{LDDef},
\begin{equation}\label{jeigen}
\partial_i \rho_\theta = \frac{R+L}{2}L_{i,\theta}^{(f_{\text{ari}})} = \frac{1}{2}\{\rho_\theta,L_{i,\theta}^{(f_{\text{ari}})}\}\;,
\end{equation}
 so that $L_{i,\theta}^{(f_{\text{ari}})}$  is also called the \emph{symmetric} logarithmic derivative (SLD)
 of $\rho_\theta$. The corresponding quantum metric is
 \begin{equation}
 [\mathcal F_Q^{(f_{\text{ari}})}(\theta)]_{ij} = \Re\, \text{tr}(\rho_\theta\, L_{i,\theta}^{(f_{\text{ari}})} L_{j,\theta}^{(f_{\text{ari}})})\;,
 \end{equation}
which is usually referred to as the \emph{quantum Fisher information} (QFI) metric and denoted simply by $\mathcal F_Q(\theta)$. It can be obtained by ``quantizing'' the \emph{Bures distance} $d_B^2$ \cite{bengtsson_geometry_2017}, in the sense that
\begin{equation}
d_B^2(\rho_\theta, \rho_{\theta+d\theta}) = \frac{1}{4}\, [\mathcal F_Q(\theta)]_{ij}\, d\theta^i d\theta^j\;,
\end{equation}
where $d_B^2(\rho,\sigma) = 2[1-\sqrt{F(\rho,\sigma)}]$ and $F(\rho,\sigma)=(\tr[\sqrt{\sqrt \rho \,\sigma\,\sqrt \rho}])^2$ is the \emph{fidelity}.
\item[\textbf{(M2)}] The operator monotone function $f_{\text{har}}= 2x/(1+x)$ corresponds to the harmonic mean, since for commuting matrix $A,B$ one has $\mathfrak{m}^{(f_{\text{har}})}(A,B)= 2 A B/(A+B)$. From Eq.~\eqref{LDDef}, one finds:
\begin{equation}
\partial_i \rho_\theta = \frac{2 LR}{L+R}L_{i,\theta}^{(f_{\text{har}})}\implies L_{i,\theta}^{(f_{\text{har}})} = \frac{1}{2}\{\rho_\theta^{-1},\partial_i \rho_\theta\}\;.
\end{equation}
The corresponding metric is
\begin{equation}
[\mathcal F_Q^{(f_{\text{har}})}]_{ij}=\Re\,\text{tr}(\partial_i \rho_\theta \partial_j \rho_\theta \rho_\theta^{-1})\;.
\end{equation}
\item[\textbf{(M3)}] The logarithmic mean corresponds to $f_{\text{log}}=(x-1)/\log x$ since, for commuting $A$ and $B$,  $\mathfrak m^{(f_\text{log})}(A,B)=(B-A)/(\log B- \log A)$. From Eq.~\eqref{LDDef}, one obtains the condition:
\begin{equation}
\partial_i \rho_\theta = \frac{L-R}{\log L -\log R} L_{i,\theta}^{(f_{\text{log}})}\implies [\log \rho_\theta,\partial_i \rho_\theta]=[\rho_\theta,\,L_{i,\theta}^{(f_{\text{log}})}]\;.
\end{equation}
One can solve for $L_{i,\theta}^{(f_{\text{log}})}$ as follows. First of all, let us recall the identity
\begin{equation}
\log \rho_\theta = \int_0^\infty \frac{dt}{1+t} - \int_0^\infty \frac{dt}{\rho_\theta  + t}\;.
\end{equation}
The commutator $[\log \rho_\theta, \partial_i \rho_\theta]$ can now be rewritten as follows:
\begin{equation}\label{logeq}
\begin{split}
  [\log \rho_\theta, \partial_i \rho_\theta] =& \int_0^\infty  dt\, \left[\partial_i \rho_\theta, \frac{1}{\rho_\theta+t}\right] \\
  =& \int_0^\infty  dt\,\left[\partial_i\frac{1}{\rho_\theta+t}, \rho_\theta \right]\\
  =& \int_0^\infty dt\, \left[\rho_\theta, \frac{1}{\rho_\theta+t} \partial_i \rho_\theta \frac{1}{\rho_\theta+t}\right]
\end{split}
\end{equation}
where we made use of the fact that, for any invertible matrix $M$, $\partial_i M^{-1} = - M^{-1} \partial_i M M^{-1}$.
From Eq.~\eqref{logeq}, $L_{i,\theta}^{(f_{\text{log}})}$ can be read-off directly, i.e.
\begin{equation}
  L_{i,\theta}^{(f_{\text{log}})} = \int_0^\infty  dt\, \frac{1}{\rho_\theta+t} \partial_i \rho_\theta \frac{1}{\rho_\theta+t}\;.
\end{equation}
The corresponding metric is the \emph{Bogoliubov-Kubo-Mori metric}:
\begin{equation}
  [\mathcal F_Q^{(f_{\text{log}})}]_{ij} = \int_0^\infty dt\,\tr\left(\partial_i \rho_\theta \frac{1}{\rho_\theta+t} \partial_j \rho_\theta \frac{1}{\rho_\theta+t}\right)\;.
\end{equation}
It can be obtained by ``quantizing'' the \emph{quantum relative entropy} $S(\rho||\sigma)$ \cite{bengtsson_geometry_2017}, in the sense that
\begin{equation}
S(\rho_{\theta}||\rho_{\theta+d\theta}) = \frac{1}{2}\, [\mathcal F_Q(\theta)^{(f_{\text{log}})}]_{ij}\, d\theta^i d\theta^j\;,
\end{equation}
where $S(\rho||\sigma)=\tr[\rho(\log \rho-\log \sigma)]$.
\end{itemize}
It is also possible to derive a closed-form expression for $\mathcal F_Q^{(f)}$, with $f$ an arbitrary operator monotone function. Notice that the superoperators $L$ and $R$ commute. Moreover, if $\rho_\theta = \sum_{k=1}^d p_k \ket{k}\bra{k}$ (where $\{\ket{k}\}_{k=1}^d$ are the normalized eigenvectors of $\rho_\theta$), then
\begin{equation}
  L \ket{k}\bra{l} = p_k \ket{k}\bra{l}\;, \qquad R \ket{k}\bra{l} = p_l \ket{k}\bra{l}\;.
\end{equation}
It follows that $\{\ket{k}\bra{l}\}_{k,l=1}^d$ is a complete system of eigenvectors for both $R$ and $L$. They are also the eigenvectors of the superoperator $\mathcal J = R f(LR^{-1})$, with eigenvalues:
\begin{equation}
  \mathcal J \ket{k}\bra{l} = p_l\, f\left(\frac{p_k}{p_l}\right)\ket{k}\bra{l}\;.
\end{equation}
Let us expand the symmetric derivative operators as
\begin{equation}\label{logDerexp}
  L_{i,\theta}^{(f)} = \sum_{k,l=1}^d \ell_{kl}^{(i)}\ket{k}\bra{l}\;.
\end{equation}
Notice that since $\rho_\theta$ is full-rank, the coefficients $\ell_{kl}^{(i)}$ completely determine $L_{i,\theta}^{(f)}$. Next, one substitutes Eq.~\eqref{logDerexp} into Eq.~\eqref{logeq} and compares terms, which leads to the conditions:
\begin{equation}
  \ell_{kl}^{(i)} = \begin{dcases}
    \frac{\partial_i p_k}{p_k}\qquad\qquad \qquad \quad\;\;\;  (k=l)\;,\\[5pt]
    \frac{p_l-p_k}{p_l\,f(p_k/p_l)} \braket{k|\partial_i l}\qquad (k\neq l)\;.
  \end{dcases}
\end{equation}
From Eq.~\eqref{qMetric} and the previous relation, one finds:
\begin{equation}
  [\mathcal F_Q^{(f)}(\theta)]_{ij} = \sum_{k=1}^d \frac{\partial_i p_k\, \partial_j p_k}{p_k} + \sum_{l\neq k} \frac{(p_l - p_k)^2}{p_l\, f(p_k/p_l)} \braket{k|\partial_i l}\braket{\partial_j l|k}\;,
\end{equation}
which is our final result.
If the statistical model is not full-rank, one can still recover all possible monotone metrics by extending the metrics of Eq.~\eqref{petzclass} via a suitable fiber bundle construction (see e.g.~\cite{petz_geometries_1996}). In particular, for a pure model $S=\{\ket{\psi_\theta}\}_{\theta\in\Theta}$, the extension of the metric $\mathcal F_Q^{(f)}$ on $S$ exists if and only if $f(0)\neq 0$, in which case it is always proportional to the Fubini-Study metric (which is in fact the unique unitarily invariant metric on pure states \cite{bengtsson_geometry_2017}). For instance, the quantum Fisher information metric evaluates to:
\begin{equation}\label{FSmetric}
  [\mathcal F_Q(\theta)]_{ij} = 4 \Re\, \left[\braket{\partial_i \psi_\theta|\partial_j \psi_\theta}+\braket{ \psi_\theta|\partial_i \psi_\theta}\braket{\psi_\theta|\partial_j \psi_\theta}\right]\;.
\end{equation}
See also Ref.~\cite{jing_quantum_2014} for a closed-form expression of $\mathcal F_Q(\theta)$ when $1<\text{rk}(\rho_\theta)<d$.
\par
In spite of the infinite number of possible metrics, Braunstein and Caves \cite{braunstein_statistical_1994} have shown that the  quantum Fisher information metric $\mathcal F_Q(\theta)$ is the only relevant one from an estimation viewpoint. This is true, at least, in the case of uniparametric models (i.e., when there is only one parameter $\theta=\theta^1$ to be estimated), to which from now on we restrict our attention (see however Rem.~\ref{multip}). 
Let us recall that a typical quantum estimation protocol is specified by a triple $(\rho_0,\mathcal M,\hat \theta^{(N)})$ and can be broken down into the following steps:
\begin{itemize}
\item[\textbf{(S1)}] \textbf{Initialization:} The statistical model $\rho_\theta$ is prepared by suitably encoding the parameter into an initial state $\rho_0$.
\item[\textbf{(S2)}] \textbf{Measurement:} A measurement $\mathcal M$ is performed, yielding an outcome $x\in\mathcal X$. When $N$ independent measurements are taken onto identically prepared systems, one obtains a sample $\boldsymbol x \in\mathcal X^{\times N}$.
\item[\textbf{(S3)}] \textbf{Data processing:} The sample $\boldsymbol x$ is processed through the estimator $\hat \theta^{(N)}$.
\end{itemize}
The problem is to optimize over each step in order to minimize a given objective function, which is generally taken to be the mean-square-error $\text{MSE}(\hat \theta^{(N)})$. Notice that, among the three steps, only \textbf{(S1)} and \textbf{(S2)} are properly quantum. Moreover, in the asymptotic limit of a large number of sample points,  optimization over \textbf{(S3)} is trivially carried out by employing an asymptotically efficient estimator. In contrast, optimization over the measurement step \textbf{(S2)} is a non-trivial task. However, as long as $N\gg 1$, minimization of $\text{MSE}(\hat \theta^{(N)})$ is equivalent to maximization of the Fisher information $\mathcal F_C(\theta)$ corresponding to the classical statistical model $p_{\theta}(x) = \tr(\rho_\theta \Pi_x)$
(with $\{\Pi_x\}_{x\in\mathcal X}$ the probability operators of a generic measurement $\mathcal M$). Therefore, the strategy usually followed is first to identify the family $\mathscr F$ of measurements which are available to the experimentalist, and then to maximize the Fisher information over all measurements $\mathcal M \in \mathscr F$. 
\par
We now introduce the family of \emph{regular} measurements.
\begin{defin}\textbf{(regular measurement)}
A measurement $\mathcal M$ is called regular if its probability operators are parameter-independent, i.e.
\begin{equation}
  \partial_\theta \Pi_x = 0\;,\qquad  \forall x \in\mathcal X\;;
\end{equation}
otherwise, the measurement is non-regular. 
\end{defin}
Braunstein and Caves have maximized the Fisher information over the family $\mathscr F_R$ of regular measurements.
\begin{thm}\label{BCthmBound} \textbf{(Braunstein-Caves \cite{braunstein_statistical_1994})}
For uniparametric model, the maximum Fisher information, optimized over the family $\mathscr F_R$, is the quantum Fisher information:
\begin{equation}
  \mathcal F_Q(\theta) = \underset{\mathcal M \in \mathscr F_R}{\text{\emph{max}}}\, \mathcal F_C(\theta)\;.
\end{equation}
\begin{proof}
For a generic measurement, the Fisher information can be written as
\begin{equation}\label{eq1BC}
  \mathcal F_C(\theta) = \sum_{x\in\mathcal X^*} \frac{[\partial_\theta \tr(\rho_\theta \Pi_x)]^2}{\tr(\rho_\theta \Pi_x)}\;,
\end{equation}
where $\mathcal X^*\eqd \{x\in\mathcal X:\, \tr(\rho_\theta \Pi_x)\neq 0\}$. Notice that, in Def.~\eqref{FIDefin}, summation is  only over those outcomes belonging to the support of $p_\theta$. In the quantum case the role of $p_\theta$ is taken by $\text{Pr}_\theta(x)=\tr(\rho_\theta \Pi_x)$, so one should exclude
outcomes $x\in \mathcal X \setminus \mathcal X^*$ for which $\text{Pr}_\theta(x)=0$. This clarification becomes irrelevant if $\rho_\theta$ is full-rank, since then $\mathcal X=\mathcal X^*$. Eq.~\eqref{eq1BC} can be manipulated as follows:
\begin{align}\label{manipul}
  \mathcal F_C(\theta) =&\,\sum_{x\in\mathcal X^*} \frac{\Re^2\,\tr(\rho_\theta L_\theta \Pi_x)}{\tr(\rho_\theta \Pi_x)}\\
  \leq & \,\sum_{x\in\mathcal X^*}\frac{|\tr(\rho_\theta L_\theta \Pi_x)|^2}{\tr(\rho_\theta \Pi_x)}\\
  =&\,\sum_{x\in\mathcal X^*}\frac{|\tr(\sqrt{\Pi_x}\sqrt{\rho_\theta}\sqrt{\rho_\theta} L_\theta \sqrt{\Pi_x})|^2}{\tr(\rho_\theta \Pi_x)}\\
  \leq & \,\sum_{x\in\mathcal X^*}\frac{\tr(\rho_\theta \Pi_x)\tr( L_\theta\rho_\theta L_\theta \Pi_x)}{\tr(\rho_\theta \Pi_x)}\\
  =&\,\sum_{x\in\mathcal X^*} \tr(L_\theta \rho_\theta L_\theta \Pi_x)\\
  \leq & \,\sum_{x\in\mathcal X}\tr(L_\theta \rho_\theta L_\theta \Pi_x)\\
  =&\,\tr(\rho_\theta L_\theta^2)
  = \mathcal F_Q(\theta)\;.
\end{align}
In the first line, we have employed the defining relation of the symmetric logarithmic derivative $\partial_\theta \rho_\theta = \{\rho_\theta, L_\theta\}/2$; in the second line, the inequality $\Re^2 z\leq |z|^2$, $\forall z \in \mathbb C$; in the fourth line, the Cauchy-Schwarz inequality; in the sixth, we have extended summation over all outcomes $\mathcal X$, noting that $\tr(L_\theta \rho_\theta L_\theta \Pi_x)\geq 0$, $\forall x\in\mathcal X$
\footnote{In fact, $L_\theta \rho_\theta L_\theta$ and $\Pi_x$ are positive-semidefinite matrices and the trace of the product of two positive semi-definite matrices is always nonnegative.}; finally, in the last line, we have made use of the completeness relation $\sum_{x\in\mathcal X}\Pi_x= \mathbb I_d$. We have thus proved that, for any regular measurement $\mathcal M\in \mathscr F_R$, $\mathcal F_C(\theta)\leq \mathcal F_Q(\theta)$. 

We will now show that there always exists a measurement saturating the previous inequality, which will establish the theorem. The above manipulations involved three separate inequalities, that to be simultaneously saturated require:
\textbf{(R1)}: $\Im\,\tr(\rho_\theta L_\theta \Pi_x) = 0$,  $\forall x \in \mathcal X^*$; \textbf{(R2)}: There exist complex numbers $\{\alpha_x\}_{x\in\mathcal X^*}$ such that $\sqrt{\rho_\theta} L_\theta \sqrt{\Pi_x} = \alpha_x \sqrt{\rho_\theta} \sqrt{\Pi_x}$; \textbf{(R3)}: $\sum_{x\in\mathcal X\setminus \mathcal X^*} \tr(L_\theta \rho_\theta L_\theta \Pi_x) =0$. 
It is easy to check that requirements \textbf{(R1)} through \textbf{(R3)} are satisfied by performing a projective measurement of the symmetric logarithmic derivative $L_\theta$. More precisely, let us remark that the defining relation $\partial_\theta \rho_\theta = \{\rho_\theta, L_\theta\}/2$ determines $L_\theta$ only on the support of $\rho_\theta$: outside the support $\text{supp}(\rho_\theta)$, $L_\theta$ may be defined in an arbitrary way, compatible with Hermiticity. The SLD  $L_\theta$ may thus be written as follows:
\begin{equation}
L_\theta = \sum_{x\in\mathcal X} \lambda_{x,\theta} \ket{\lambda_{x,\theta}}\bra{\lambda_{x,\theta}} \;,
\end{equation}
where $\{\ket{\lambda_{x,\theta}}\}_{x\in\mathcal X\setminus \mathcal X^*}$ are chosen arbitrarily so as to give rise to an orthonormal basis. The eigenvectors and eigenvalues of $L_\theta$ are, in general, parameter-dependent. Then, if $\theta^*$ is the actual value of the parameter to be estimated, the optimal measurement is described by
\begin{equation}\label{conditionBC}
\Pi_x^{(opt)} = \ket{\lambda_{x,\theta}}\bra{\lambda_{x,\theta}}\,\big|_{\theta=\theta^*}\;,\qquad \forall x \in \mathcal X\;,
\end{equation} 
i.e.~the corresponding Fisher information satisfies $\mathcal F_C(\theta^*) = \mathcal F_Q(\theta^*)$. Notice that, for each $\theta^* \in \Theta$, there is a \emph{different} optimal measurement: it is not required to engineer the measurement so that it satisfies Eq.~\eqref{conditionBC} for any possible value of $\theta^*$. Such a measurement would instead have probability operators $\ket{\lambda_{x,\theta}}\bra{\lambda_{x,\theta}}$ and would be non-regular. However, implementing the optimal measurement \emph{does} require to know the value of $\theta^*$ for the problem at hand, which is a priori unknown. The obstacle is overcome by employing an \emph{adaptive procedure}, which involves constructing a sequence of estimates $\{\theta^*_n\}$ such that $\theta^*_n \to \theta^*$ and modifying the implemented measurement at each step  so as to match condition \eqref{conditionBC}. See e.g.~Ref.~\cite{nagaoka_asymptotically_1988} for more details.  
\end{proof}
\end{thm}
\begin{rem}\label{multip}
One may generalize Thm.~\ref{BCthmBound} to the multiparameter case. The quantum Fisher information $\mathcal F_Q(\theta)$ can be proven to be the least monotone metric such that $\mathcal F_Q(\theta) -\mathcal F_C(\theta)$ is positive semi-definite for any regular measurement. However, equality is not in general attainable, unless the commutativity condition $\tr(\rho_\theta [L_{i,\theta},L_{j,\theta}])=0$ is satisfied $\forall i,j\in\{1,\dots,m\}$ \cite{ragy_compatibility_2016,matsumoto_new_2002}. A widely employed solution \cite{gill_state_2000} is to regularize the problem, by changing the objective function to $\tr[C\cdot \text{\emph{Cov}}(\theta)]$ (where $C$ is a positive-definite diagonal matrix assigning different weights to different parameters). However, for this problem, the QFI metric is no longer necessarily the one providing the tightest bound \cite{yuen_multiple-parameter_1973}. 
\end{rem}
With some caveats, the quantum Fisher information therefore sets the ultimate asymptotic sensitivity bound in uniparametric problems.
\begin{thm}\textbf{\emph{(quantum Cram\'er-Rao)}}
For regular models and any uniparametric estimation protocol $(\rho_0,\mathcal M,\hat \theta^{(N)})$, where  $\mathcal M \in \mathscr F_R$ and the estimator $\hat \theta^{(N)}$ is unbiased, the following inequality holds:
\begin{equation}\label{QCRb}
\text{\emph{Var}}(\hat \theta^{(N)})  \geq \frac{1}{N\cdot \mathcal F_Q(\theta)}\;.
\end{equation} 
\end{thm}

As in the classical case, the bound \ref{QCRb} is saturable only for a few special statistical models (see Ref.~\cite{barndorff-nielsen_quantum_2003} for a precise statement). In contrast, in the asymptotic limit $N\gg 1$, one has that, for any regular measurement and any consistent estimator, 
\begin{equation}
\lim_{N\to \infty} N\cdot \text{\emph{Var}}(\hat \theta^{(N)}) \geq \frac{1}{ \mathcal F_Q(\theta)}\;.
\end{equation}
Equality can be achieved by resorting to the optimal measurement of Eq.~\eqref{conditionBC} and to an asymptotically efficient estimator. 
The last logical step is to maximize the QFI over the choice of the initial state $\rho_0$. To this end, the following \emph{extended convexity} property is going to be useful.
\begin{prop}\label{extConv}
Given a quantum statistical model $S = \{\rho_\theta\}_{\theta \in \Theta}$, where each $\rho_\theta$ is written as a convex superposition of the form $\rho_\theta = \sum_i \lambda_{i,\theta}\, \rho_{i,\theta}$, the quantum Fisher information satisfies the inequality:
\begin{equation}
\mathcal F_Q[\rho_\theta] \leq \sum_{i} \lambda_{i,\theta} \mathcal F_Q[\rho_{i,\theta}] + \mathcal F_C[\{\lambda_{i,\theta}\}]\;.
\end{equation}
\end{prop}
The terms in square brackets specify the statistical models on which the (quantum)  Fisher information is computed. From Prop.~\ref{extConv}, assuming that the system is prepared in the parameter-independent state $\rho_0 = \sum_i \lambda_i \rho_i$ and that the parameter is encoded via a channel $\mathcal E_\theta$, one has 
\begin{equation}
\mathcal F_Q[\rho_\theta] \leq \sum_i \lambda_i \mathcal F_Q[\mathcal E_\theta(\rho_i)]\;;
\end{equation}
notice that the classical term $\mathcal F_C[\{\lambda_i\}]$ vanishes since $\partial_\theta \lambda_i=0$. It follows that the QFI achieves its maximum on the set of pure states. It is not possible, in general, to further determine  the optimal preparation, with the significant exception of unitary models. 

Let us assume that $\rho_0 = \ket{\psi_0}\bra{\psi_0}$ and the encoding is provided by the unitary channel associated to $U_t=\text{exp}(-itH_\theta)$. Then, substituting into Eq.~\eqref{FSmetric}, one obtains
\begin{equation}\label{qfiPure}
\mathcal F_Q(\theta) = 4 [\braket{\psi_\theta|\mathfrak g_{\theta}^2[U_t] |\psi_\theta}-(\braket{\psi_\theta|\mathfrak g_{\theta}[U_t] |\psi_\theta})^2]\;,
\end{equation}
where $\ket{\psi_\theta} = U_t \ket{\psi_0}$ and $\mathfrak g_{\theta}[U_t]\eqd i\partial_\theta U_t U_t^\dagger$ is the local generator of $U_t$. Eq.~\eqref{qfiPure} may be rewritten as
\begin{equation}
\mathcal F_Q(\theta) = 4\, \text{Var}_{U_t \ket{\psi_0}}\left[\mathfrak g_\theta[U_t]\right] = 4\, \text{Var}_{\ket{\psi_0}}[U_t^\dagger\, \mathfrak g_\theta[U_t]\, U_t]\;,
\end{equation}
where $\text{Var}_{\ket{\psi}}[O]$ is by definition the variance of the operator $O$ over a state $\ket{\psi}$. Let us recall that, by Popoviciu's inequality \cite{popoviciu_tiberiu_sur_1935}, for any random variable $\mathbb Y$,
	\begin{equation}\label{popineq}
	 \text{Var}(\mathbb Y)\leq \frac{(Y-y)^2}{4}\;,
	\end{equation}
where $Y$ (\emph{resp}. $y$) is the maximum (\emph{resp}. minimum) value of $\mathbb Y$ and equality holds when $\mathbb Y$ is equally distributed over the two values $Y$ and $y$. Let us also introduce the following standard notation for the eigenvalues of a matrix $M\in \mathsf{Her}_d(\mathbb C)$: if we denote the $d$ real eigenvalues of $M$ by $\text{spec}(M) = \{\lambda_1(M),\dots,\lambda_d(M)\}$ (ordered non-decreasingly, i.e.~$\lambda_1(M)\geq \dots\geq \lambda_d(M)$), it then follows that 
\begin{align*}
\mathcal F_Q(\theta)  \leq [\lambda_1(U_t^\dagger \mathfrak g_\theta[U_t]U_t)- \lambda_d(U_t^\dagger \mathfrak g_\theta[U_t]U_t)]^2 & = [\lambda_1( \mathfrak g_\theta[U_t])- \lambda_d( \mathfrak g_\theta[U_t])]^2 \\ & = [\sigma(\mathfrak g_\theta[U_t])]^2\;,
\end{align*}
where the \emph{spectral gap} of a matrix $M\in \mathsf{Her}_d(\mathbb C)$ is defined as $\sigma[M]\eqd \lambda_1(M)-\lambda_d(M)$. The equal sign 
is achieved by preapring a (any) balanced superposition of the extremal eigenvectors of the generator. Overall, we may summarize the result by 
the following proposition.
\begin{prop}
Given the unitary model $\{\rho_\theta\}_{\theta\in\Theta}$, with $\rho_\theta = U_t \rho_0 U_t^\dagger$ and $U_t = \text{exp}(-itH_\theta)$, one has
\begin{equation}
\underset{\rho_0}{\text{\emph{max}}}\; \mathcal F_Q[U_t \rho_0 U_t^\dagger] = [\sigma(\mathfrak g_\theta[U_t])]^2\;.
\end{equation}
The maximum is reached upon setting $\rho_0 = \ket{\psi_0^{(opt)}} \bra{\psi_0^{(opt)}}$, where $\ket{\psi_0^{(opt)}}$ is  a balanced superposition of the extremal eigenvectors of the generator $\mathfrak g_\theta[U_t]$:
\begin{equation}
\ket{\psi_0^{(opt)}} = \frac{1}{\sqrt 2} \left(\ket{\lambda_1(\mathfrak g_\theta[U_t])} + e^{i\phi} \ket{\lambda_d(\mathfrak g_\theta[U_t])}\right)\;,\qquad \phi \in \mathbb R\;.
\end{equation}
\end{prop}
\section{Non-regular measurements and parameter estimation beyond the quantum Cram\'er-Rao theorem}
\label{pebcr}
Let us now extend the theory of quantum parameter estimation, by enlarging the class of measurements under consideration to non-regular measurements, i.e.~measurements carrying an intrinsic dependence on the unknown value of the parameter. Such measurements will be shown to lead to an improvement of the achievable precision, beyond the bound encoded by the quantum Cram\'er-Rao theorem \cite{seveso_quantum_2017,seveso_estimation_2017}. 
\par
A measurement $\mathcal M_\theta$ is said to be \emph{non-regular} if its probability operators $\{\Pi_{x,\theta}\}_{x\in\mathcal X}$ are parameter-dependent. Since non-regular measurements, by  definition, do not belong to the family $\mathscr F_R$ over which the Fisher information was optimized in Thm.~\ref{BCthmBound}, they might outperform the optimal Braunstein-Caves measurement. Explicitly, their Fisher information $\mathcal F_C(\theta)$ reads 
 \begin{align}
\mathcal F_C(\theta) =\sum_{x\in\mathcal X^*}\!\frac{\Re^2 \tr(\rho_\theta L_\theta \Pi_{x,\theta})}{\tr(\rho_\theta \Pi_{x,\theta})} & + \sum_{x\in\mathcal X^*}\!\frac{[\tr(\rho_\theta\,  \partial_\theta \Pi_{x,\theta})]^2}{\tr(\rho_\theta \Pi_{x,\theta})} \notag \\ & + 2 \sum_{x\in\mathcal X^*}\!\frac{ \Re \tr(\rho_\theta L_\theta \Pi_{x,\theta})\tr(\rho_\theta\, \partial_\theta \Pi_{x,\theta})}{\tr(\rho_\theta \Pi_{x,\theta})}\;.
 \end{align}
The first term on the RHS is the same that appears on the first line of Eq.~\eqref{manipul} and that is bounded from above by the QFI, but there are also two additional contributions. In general, they will have an important effect on the achievable sensitivity (though they are not always positive, so a precision enhancement is not guaranteed).

It is not immediately clear how to implement non-regular measurements. Seemingly, one would need to know beforehand the true value of the parameter. The same could be said of the statistical model $\rho_\theta$ but, in the latter case, the true value of the parameter is encoded into the initial state, e.g.~by making use of the time-evolution of the system as a resource. In the same way, a non-regular measurement requires the parameter to be suitably encoded into its probability operators. We now describe two scenarios where this is possible. 
\subsection{Measurement models with parameter-dependent interactions}
\begin{figure}[h!]
\centering
\includegraphics[width=0.99\textwidth]{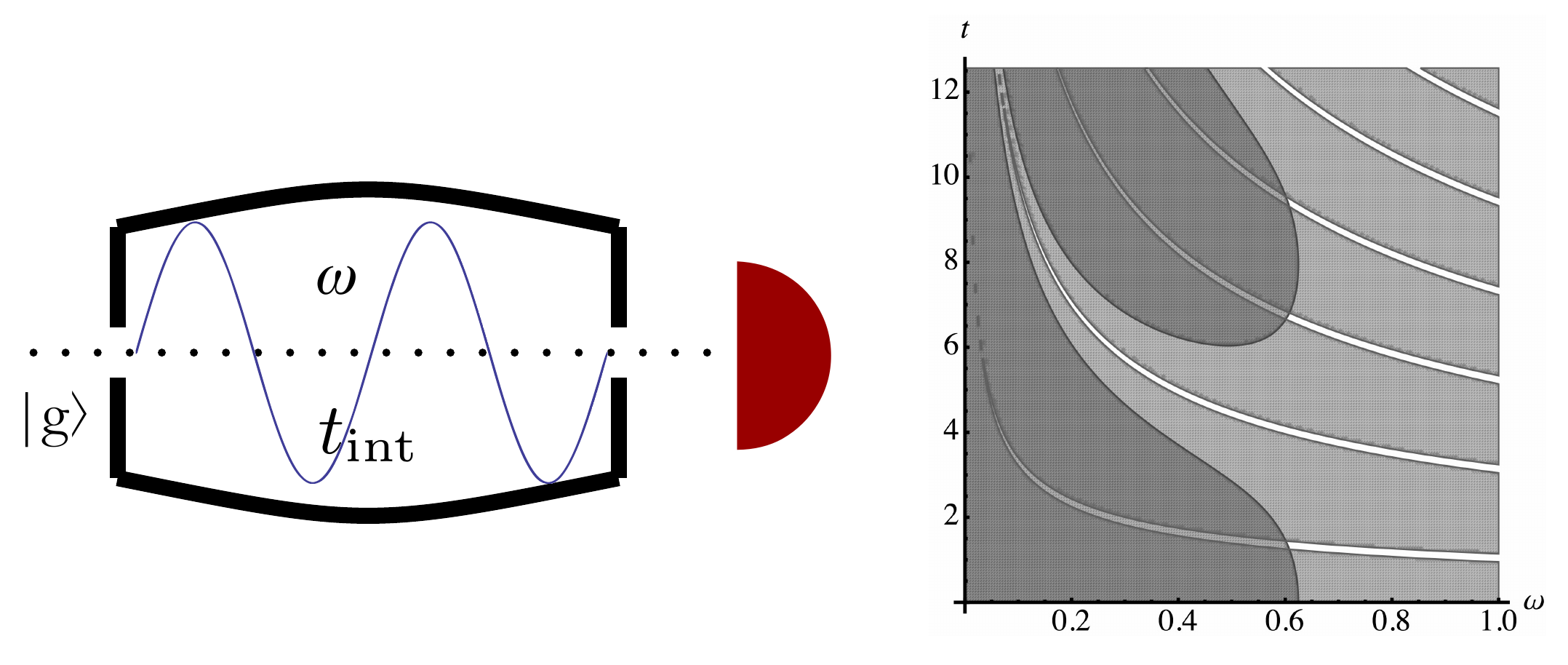}
\caption{Ancilla-assisted estimation of the frequency of a bosonic mode in a cavity. A non-regular measurement can be engineered by coupling the bosonic mode 
to a two-level atom, which is initially in its ground state $\ket g$, and by measuring whether the atom has been excited or not after an interaction time $t$. In the right panel, we show the regions in the $\omega-t$ plane where the ratio $\gamma = \mathcal F_C(\omega)/\mathcal F_Q(\omega)$ is larger than one. The dark-gray region is for $\Omega = 0.5 \sqrt{\omega}$ and 
the light-gray one for $\Omega = 1.5 \sqrt{\omega}$.}\label{f:jc}
\end{figure}

Let us model a non-regular measurement as in Sect.~\ref{qmth}, by specifying the interaction between the system and the apparatus. The total Hamiltonian is $H_T = H_{\theta} + H_A + H_{I,\theta}$, where we assume that the free Hamiltonian $H_A$ of the apparatus does not depend on the parameter, but the coupling term $H_{I,\theta}$ does. We also assume that the duration 
of the measurement $t_{int}$ is short and the interaction is strong, such 
that the free evolution of the two systems may be neglected, i.e. the 
time-evolution operator during the measurement process may be written as 
$U_{t} \sim \text{exp}(-it H_{I,\theta})$. If the apparatus is prepared in a reference state $\ket{\phi}$ and a projective measurement $\{P_x\}_{x\in\mathcal X}$ is made on the ancilla after a time $t_{int}$, the resulting probability operators read
\be
\Pi_{x,\theta} = \braket{\phi| e^{it H_{I,\theta}}\, \mathbb I_d \otimes P_x \, e^{-it H_{I,\theta}}|\phi}
\ee
and are, in general, parameter-dependent. 
A simple example of this scenario is provided by the estimation of the frequency of a bosonic mode, see a schematic diagram in Fig. \ref{f:jc}.
\par\noindent
The parameter to be estimated is the frequency $\omega$ of a bosonic mode 
in a cavity. The system's Hamiltonian is $H_\omega= \omega(a^\dagger a +1/2)$, the initial state is chosen as $\ket{\psi_0}=\alpha_0 \ket{0} + \alpha_1 \ket{1}$ and the statistical model at time $t$ is $\ket{\psi_\omega}\eqd U_t \ket{\psi_0} = \alpha_0 e^{-i\omega t/2} \ket{0}+\alpha_1 e^{-3i\omega t/2}  \ket{1}$, where $U_t \eqd \text{\emph{exp}}(-it H_\omega)$. The QFI may be written as $$\mathcal F_Q(\omega)=4t^2 |\alpha_0|^2 |\alpha_1|^2\,,$$
which is the maximum information extractable via regular measurements.
A non-regular measurement can be engineered by coupling the bosonic mode 
to a two-level atom, which is initially in its ground state $\ket g$, and by measuring whether the atom has been excited or not after an interaction time $t_{int}$. The interaction Hamiltonian is of the Jaynes-Cummings type $H_{I}= \Omega(a^\dagger \sigma_{-}+a \sigma_{+})$, where $\Omega \eqd d\sqrt{\omega/2\epsilon_0 V}$, $d \eqd \vec {\epsilon}\cdot \bra{e} \vec{d} \ket{g}$, $\vec{\epsilon}$ is the photon polarization, $\epsilon_0$ is the dielectric constant, $V$ the volume of the cavity, $\vec{d}$ the dipole operator, $\ket{g}$ the atom's ground state, $\ket{e}$ the excited state, $\sigma_+\eqd \ket{e}\bra{g}$ and $\sigma_{-}\eqd \ket{g}\bra{e}$. Notice that $\Omega = \kappa \sqrt{\omega}$ with $\kappa=d/V\sqrt{2\epsilon_0}$, such that the interaction Hamiltonian is parameter-dependent. Explicitly, the evolution operator $U_t$ during the measurement process is
\begin{equation}
  U_t = U_{gg} \ket{g}\bra{g}  +  U_{ge}  \ket{g}\bra{e} + U_{eg} \ket{e}\bra{g} + U_{ee} \ket{e}\bra{e}\;,
\end{equation}
where, letting $N \eqd a^\dagger a$ denote the number operator for the radiation field, we have defined 
\be
\begin{split}
& U_{gg} \eqd\cos(\Omega t\sqrt{N}) \;,\qquad \qquad \quad \; U_{ge} \eqd -i\frac{\sin(\Omega t\sqrt{N})}{\sqrt{N}} a^\dagger \;,\\
& U_{eg} \eqd -i\frac{\sin(\Omega t\sqrt{1+N})}{\sqrt{1+N}} a \;, \qquad U_{ee} \eqd \cos(\Omega t\sqrt{1+N})\;.
\end{split}
\ee
By convention, the outcome $0$ is obtained if the atom is measured in the ground state and the outcome $1$ if measured in the excited state. 
From Eq.~\eqref{mmPOVM}, the measurement operators and the corresponding probability operators are given by 
\begin{align}
  M_{0,\omega} &= \bra{g}U_{t}\ket{g}=\cos(\Omega t \sqrt{N})
  \;,\quad\! 
  M_{1,\omega} =\bra{e}U_{t}\ket{g} = -i\frac{\sin(\Omega t\sqrt{1+N})}{\sqrt{1+N}} a\,, \\
\Pi_{0,\omega}& =\cos^2(\Omega t \sqrt{N}) \;,\quad\! 
\Pi_{1,\omega} = \sin^2(\Omega t \sqrt{N})\,.
\end{align}
They depend on the parameter $\omega$ via the coupling constant $\Omega$. The  Fisher information is then given by
\begin{equation}
\mathcal F_C(\omega) = \left(\frac{\Omega t}{\omega}\right)^2\,\frac{|\alpha_1|^2 \cos^2(\Omega t)}{1- |\alpha_1|^2\,\sin^2(\Omega t)}\;,
\end{equation}
which is not necessarily bounded from above by the QFI. For instance, if the system is initially prepared in the excited state, then the QFI vanishes (there is no regular measurement 
that can estimate the parameter with finite precision), but 
$\mathcal F_C(\omega) = (\Omega t /\omega)^2$. More generally, for small values of $\alpha_0$ we have a diverging ratio
$\mathcal F_C(\omega)/\mathcal F_Q(\omega) \simeq \Omega^2 / (4 \omega^2 |\alpha_0|^2)$, and we have $\mathcal F_C(\omega)/\mathcal F_Q(\omega) >1$ for values of $\alpha_0$ satisfying the condition
\be
|\alpha_0|^2 < \frac{\sqrt{1+ \frac{\Omega^2}{\omega^2} \tan^2 (\Omega t) }-1}{2 \tan^2 (\Omega t)} 
\,.
\ee
In the right panel of Fig. \ref{f:jc} we show the regions in the $\omega-t$ plane where the ratio $\gamma = \mathcal F_C(\omega)/\mathcal F_Q(\omega)$ is larger than one. The dark region is for $\Omega = 0.5 \sqrt{\omega}$ and 
the light one for $\Omega = 1.5 \sqrt{\omega}$.
\subsection{Energy measurements of non-linear Hamiltonians}
\begin{figure}[h!]
\centering
\includegraphics[width=0.99\textwidth]{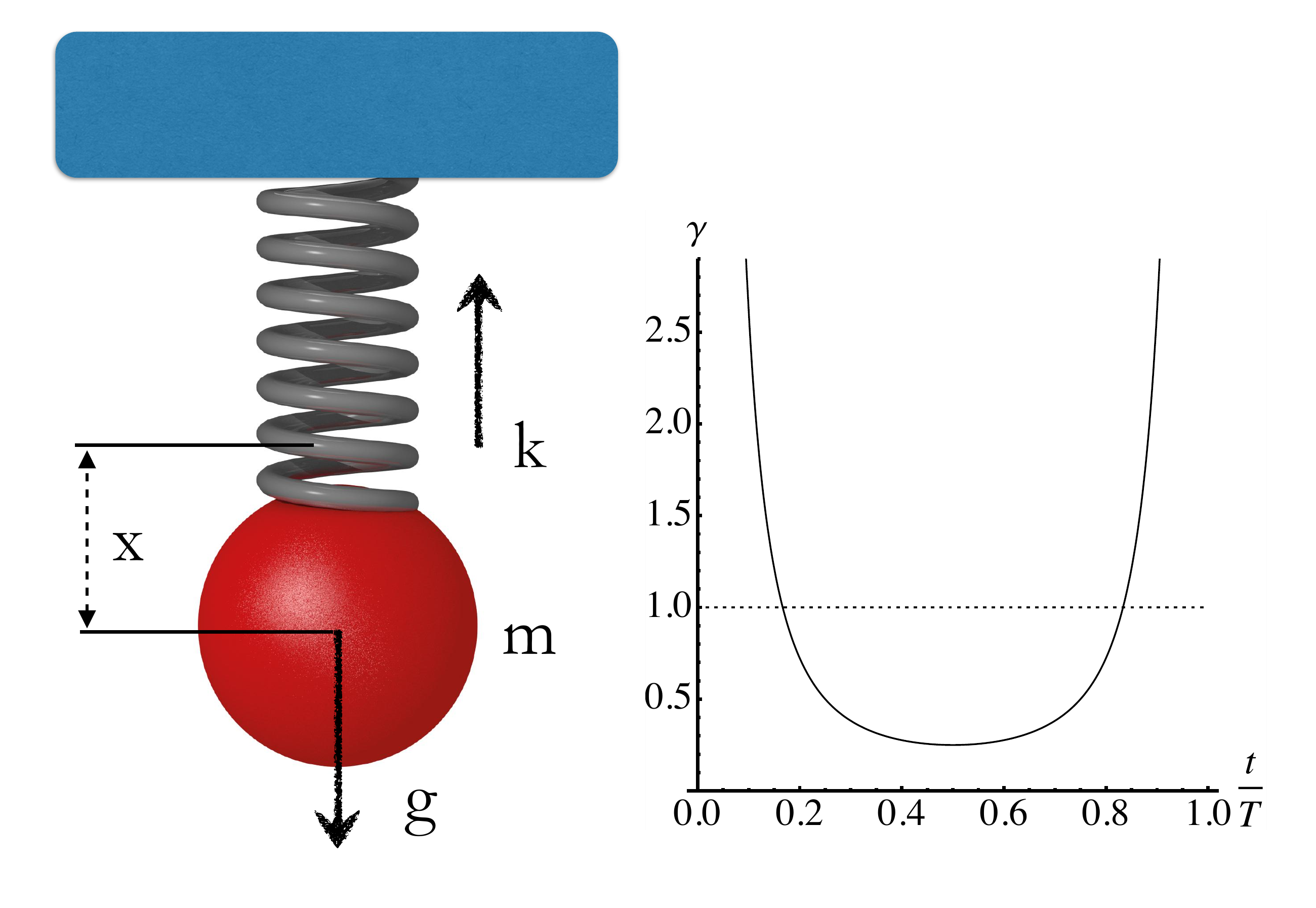}
\caption{Estimation of the strength $g$ of a uniform gravitational field using
a (quantum) mechanical oscillator as a probing system.  The mass of the oscillator is $m$, $k$ its elastic constant, and $x$ denotes the vertical displacement of the oscillator from equilibrium. In the right panel we show the ratio $\gamma=\mathcal F_C(g)/\mathcal F_Q(g)$ as a function of $t/T$, $T=2\pi/\omega$ being the period of the oscillator.}
\label{f:ho}
\end{figure}

If the Hamiltonian $H_\theta$ depends on the parameter $\theta$ in a non-linear way (i.e.~it is not of the form $H_\theta = \theta G$), its eigenstates $\{\ket{\xi_{j,\theta}}\}_{j=0}^{d-1}$ are in general parameter-dependent. An energy measurement corresponds to the projective probability operators $\Pi_{\xi_j,\theta} = \ket{\xi_{j,\theta}} \bra{\xi_{j,\theta}}$, thus the measurement is non-regular. As an example, let us consider the estimation
of the strength $g$ of a uniform gravitational field. The probing system is a mechanical oscillator, with Hamiltonian $H_g = -\partial_x^2/2m +k x^2/2+mg x$, where $m$ is the mass of the oscillator, $k$ its elastic constant and $x$ denotes the vertical displacement of the oscillator from equilibrium, see Fig. \ref{f:ho}. 
\par\noindent
The energy eigenstates have the following wavefunctions:
\begin{equation}
\psi_{j}=\left(\frac{m \omega}{\pi}\right)^{1/4}\frac{1}{\sqrt{2^j\,j!}}\,H_j(\varkappa+\varkappa_g)\,e^{-(\varkappa+\varkappa_g)^2/2}\;,
\end{equation}
where $j\in\mathbb N_0$, $H_j$ is the $j^{\text{th}}$ Hermite polynomial, $\omega\eqd \sqrt{k/m}$, $\varkappa$ is the dimensionless coordinate $\varkappa \eqd x/\ell$, $\ell$ is the characteristic length of the oscillator $\ell \eqd 1/\sqrt{m\omega}$ and $\varkappa_g \eqd mg/k\ell$. The corresponding eigenvalues are $\xi_{j,g}=\omega(j + 1/2)- mg^2/2\omega^2$. At time $t=0$, the oscillator is cooled to its ground state $\psi_{0}$; it is henceforth mechanically displaced from its equilibrium point by a distance $\delta x$, so that the initial state is
\begin{equation} 
\psi(x,0)=\left(\frac{m \omega}{\pi}\right)^{1/4}\,e^{-(\varkappa+\varkappa_\delta)^2/2}\;,\qquad \varkappa_\delta\eqd \delta x/\ell\;.
\end{equation} 
At the generic time $t$, the wavefunction of the oscillator reads
\begin{equation} 
\psi(x,t)=\left(\frac{m \omega}{\pi}\right)^{1/4}\, e^{-i\omega t(1- \varkappa_{g}^2)/2}\, e^{-(\varkappa+\varkappa_{g})^2/2}\,\text{\emph{exp}}\left[\Phi_g\right]\;,
\end{equation}
where
\be\label{smExEn}
\Phi_g = -e^{-i \omega t}\left(\frac{(\varkappa_\delta-\varkappa_g)^2}{2}\cos{\omega t} +(\varkappa_\delta-\varkappa_g)(\varkappa+\varkappa_g)\right)\;.
\ee
The computation of the QFI for the statistical model of Eq.~\eqref{smExEn} can be carried out straightforwardly (see Ref.~\cite{seveso_quantum_2017} for details); the final result is
\be 
\mathcal F_Q(g) =  \frac{8m}{\omega^3} \sin^2 \left(\frac{\omega t}{2}\right)\;.
\ee
It should be compared with the Fisher information $\mathcal F_C(g)$ corresponding to an energy measurement, which is independent on time and given by $\mathcal F_C(g) = 2 m/\omega^3$. Notice that $\mathcal F_C(g)$ exceeds the QFI for certain values of the interrogation time $t$, i.e. whenever  
$|\sin (\omega t/2)| < 1/2$, e.g. for $\omega t < \pi/3$ (see the left panel of Fig. \ref{f:ho}, where we show the ratio $\gamma=\mathcal F_C(g)/\mathcal F_Q(g)$ as a function of $t/T$, $T=2\pi/\omega$ being the period of the oscillator).

\section{Non-regular estimation of general Hamiltonian parameters}\label{s:nonr}
In this section, we further study non-regular estimation protocols based on energy measurements of non-linear Hamiltonians. The plan is to introduce a family of measurements that are non-regular and have a clear-cut physical interpretation;  to maximize the Fisher information over such a family; to  identify the best-performing measurement and, finally, to compare it with the optimal Braunstein-Caves measurement. 

\subsection{Controlled energy measurements}
\begin{figure}[h!]
\centering
\includegraphics[width=0.99\textwidth]{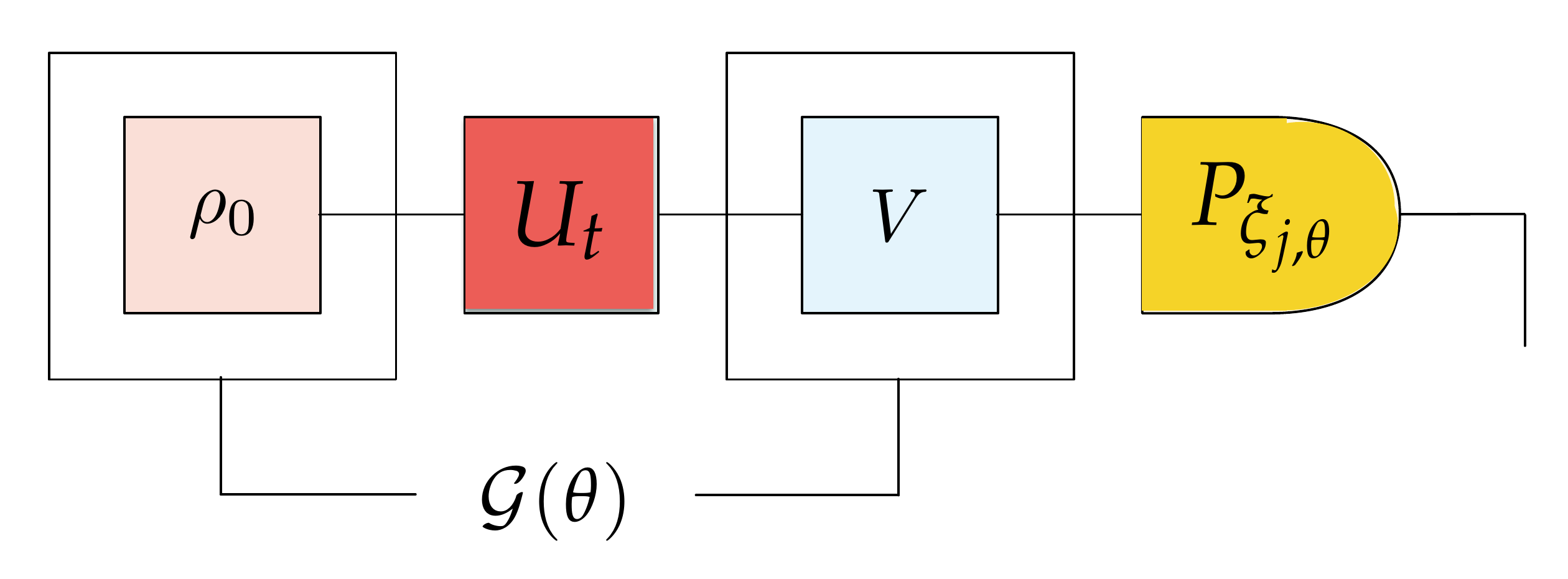}
\caption{Schematic diagram of an estimation strategy based 
on a controlled energy measurement. The optimal performance is 
quantified by ${\cal G}(\theta)$, see Eq.(\ref{cefid}), which 
is optimized over both the preparation stage and the unitary 
control. The different stages of the schemes correspond
to preparation ($\rho_0$), encoding (${\cal U}_t$), control ($V$), and 
energy measurement ($P_{\xi_{j,\theta}})$.
\label{ctrlE}}
\end{figure}

Let us consider a projective measurement of $H_\theta$, with $\theta$  a general Hamiltonian parameter. It is assumed that $H_\theta$ has eigenvalues $\xi_{j,\theta} = \lambda_{d-j}(H_\theta)$. With no significant loss of generality, the spectrum is taken to be non-degenerate. The probability of each measurement outcome is
\be\label{statEM}
\text{Pr}_\theta (\xi_{j,\theta}) = \tr(\rho_\theta P_{\xi_{j,\theta}}) = \braket{\xi_{j,\theta}|\rho_0|\xi_{j,\theta}}\;,
\ee
where $\rho_\theta = U_t \rho_0 U_t^\dagger$, $U_t = \text{exp}(-itH_\theta)$ and $P_{\xi_{j,\theta}} = \ket{\xi_{j,\theta}}\bra{\xi_{j,\theta}}$. 
The corresponding sample space $\mathcal X_\theta = \{\xi_{j,\theta}\}_{j=0}^{d-1}$ is, in general, parameter-dependent. This is a significant complication, since there is no established theory for statistical models with parameter-dependent sample spaces. In fact, if the sample space is allowed to depend on $\theta$, the proof of the classical Cram\'er-Rao theorem, Thm.~\ref{BCthmBound}, breaks down. In some cases, it is even possible to construct unbiased estimators having vanishing variance \cite{akahira_non-regular_2012}. To exclude such pathological situations, we assume in the following that either the eigenstates of $H_\theta$ are parameter-dependent, but not its eigenvalues; or that the outcomes of an energy measurement are processed via a suitable statistic $Y:\mathcal X_\theta \to \mathcal Y$, where $\mathcal Y$ is a conventional parameter-independent sample space. Within these assumptions, estimators having vanishing variance no longer occur and the 
Fisher information is again providing the relevant bounds.
Let us now consider a specific family of non-regular measurements, which wer refer to as \emph{controlled energy measurements}. They are obtained by first applying a unitary control $V\in U(d)$ and then performing a projective
energy measurement (see Fig. \ref{ctrlE}). 
\begin{defin}\textbf{(controlled energy measurement)} A controlled energy measurement $\mathcal M^{(V)}_{\theta}$ has sample space $\mathcal Y=\{\zeta_j \eqd Y(\xi_{j,\theta})\}_{j=0}^{d-1}$ and probability operators $\{\Pi_{\zeta_j}\}_{j=0}^{d-1}$, where $\Pi_{\zeta_j}\eqd V^\dagger P_{\xi_{j,\theta}} V$, $V\in U(d)$ is a unitary parameter-independent control and $P_{\xi_{j,\theta}}$ is the projector over the $j^{\text{th}}$ energy eigenstate of $H_\theta$.
\end{defin}
The Fisher information of a controlled energy measurement $\mathcal M^{(V)}_{\theta}$ is denoted by $\mathcal F_C^{(V)}(\theta)$. Let us remark that an energy measurement corresponds to the choice $V=\mathbb I_d$. Its probability measure (see Eq.~\eqref{statEM}) is $t$-independent, which implies that also the Fisher information does not depend on $t$. In contrast, the QFI generically grows quadratically with $t$ \cite{pang_quantum_2014}. Therefore, for sufficiently long times, an energy measurement can never outperform the optimal Braunstein-Caves measurement. If, on the other hand, a unitary 
control is applied before performing the measurement, then the Fisher information $\mathcal F_C^{(V)}(\theta)$ may grow again like $t^2$ 
and, in fact, it may even outperform the optimal Braunstein-Caves 
measurement at any $t$, as it will be discussed in the following. 
If an experimentalist is allowed to implement arbitrary controlled energy measurements, the maximum Fisher information she can extract is 
\be\label{cefid}
\mathcal G(\theta) \eqd \underset{\rho_0}{\text{max}}\,\underset{V\in U(d)}{\text{max}} \;\mathcal F_C^{(V)}[\rho_\theta]\;.
\ee
Compared with regular measurements, an enhancement is achievable if and only if $\mathcal G(\theta) > [\sigma(\mathfrak g_\theta[U_t])]^2$. However, computing $\mathcal G(\theta)$ directly from its definition is a non-trivial task. In the following section, a closed-form formula for $\mathcal G(\theta)$ is derived under the assumption that the Hamiltonian $H_\theta$ satisfies a rather general condition. 
\subsection{A tight achievable bound for the precision of controlled energy measurements}
For a generic controlled energy measurement $\mathcal M^{(V)}_{\theta}$, the probability of the outcome $\zeta_j$ is
\be\label{proby}
\text{Pr}_\theta(\zeta_j) = \tr(\rho_\theta V^\dagger P_{\xi_{j,\theta}} V)\;.
\ee
Let us denote by $\{\ket{j}\}_{j=0}^{d-1}$ the computational basis on the Hilbert space $\mathcal H$ of the system. The two orthonormal basis $\{\ket{j}\}_{j=0}^{d-1}$ and $\{\ket{\xi_{j,\theta}}\}_{j=0}^{d-1}$ are connected by a unitary transformation, denoted by $S \in U(d)$, such that $\ket{j}=S \ket{\xi_{j,\theta}}$.  Explicitly, the matrix elements of $S$ are $\braket{j|S|k} = \braket{\xi_{j,\theta}|k}$. Notice that, for a general Hamiltonian parameter, the matrix $S$ is $\theta$-dependent and that $S$ reduces $H_\theta$ to diagonal form, i.e.~$S H_\theta S^\dagger = \text{diag}(\xi_{0,\theta},\,\dots,\,\xi_{d-1,\theta})$. One may thus rewrite Eq.~\eqref{proby} as follows,
\be
\text{Pr}_\theta(\zeta_j) = \tr\left[(S V U_t) \rho_0 (S V U_t)^\dagger P_j \right] = \tr\big(\tilde{U}^{(V)} \rho_0 {\tilde{U}^{(V)\dagger}} P_j\big)\;,
\ee
where $P_j \eqd \ket{j}\bra{j}$ and all dependence on $\theta$ has been collected into the unitary matrix $\tilde{U}^{(V)} \eqd S V U_t$. Formally, a controlled energy measurement on the model $\rho_\theta$ is equivalent to a projective measurement in the computational basis on the model $\rho^{(V)}_\theta \eqd  \tilde{U}^{(V)} \rho_0 {\tilde{U}^{(V)\dagger}}$. The Fisher information corresponding to $\mathcal M^{(V)}_{\theta}$ can thus be written as
\be\label{FIcem}
\mathcal F_C^{(V)}(\theta) = \sum_{j\in \mathcal J^*} \frac{[\partial_\theta \tr(\rho^{(V)}_\theta P_j)]^2}{\tr(\rho^{(V)}_\theta P_j)}\;,
\ee
where $\mathcal J^*$ is the subset of $\mathcal J \eqd \{0,\dots,d-1\}$ such that $j \in\mathcal J^*$ if and only if $\text{Pr}_\theta(\zeta_j)\neq 0$. The task is to maximize the RHS of Eq.~\eqref{FIcem} over the unitary group $U(d)$ of available controls $V$ and over the initial preparation $\rho_0$. 

\begin{thm}\label{thm1BCR}
The maximum Fisher information $\mathcal G(\theta)$ that can be extracted via controlled energy measurements satisfies the inequality
\be\label{thmBCR1}
\mathcal G(\theta) \leq\left[\sigma(\mathfrak g_\theta[U_t]) + \sigma(\mathfrak g_\theta[S])\right]^2\;,
\ee 
where $U_t=\text{exp}(-it H_\theta)$ is the unitary encoding, $S$ is the similarity transformation diagonalizing $H_\theta$, $\mathfrak g_\theta[U_t]$ (\emph{resp.}, $\mathfrak g_\theta[S]$) is the generator of $U_t$ (\emph{resp.}, $S$), i.e.
\be
\mathfrak g_\theta[U_t] = i \partial_\theta U_t U_t^\dagger\;,\qquad \mathfrak g_\theta[S] = i \partial_\theta S S^\dagger\;,
\ee
and $\sigma(M)$ denotes the spectral gap of a matrix $M\in \mathsf{Her}_d(\mathbb  C)$. 
\begin{proof}
The Fisher information for $\mathcal M^{(V)}_{\theta}$ is given by Eq.~\eqref{FIcem}. Introducing the symmetric logarithmic derivative $L_\theta^{(V)}$ of $\rho^{(V)}_\theta$,
\be
\mathcal F_C^{(V)}(\theta) = \sum_{j\in \mathcal J^*} \frac{\Re^2\tr(\rho^{(V)}_\theta L^{(V)}_\theta P_j)}{\tr(\rho^{(V)}_\theta P_j)}\;.
\ee
Using the inequality $\Re z \leq |z|$, $\forall z \in \mathbb C$, and then the Cauchy-Schwarz inequality, the numerator can be bounded as follows,
\be\label{step1ineq12}
\begin{split}
\Re^2\tr(\rho^{(V)}_\theta L^{(V)}_\theta P_j) \leq &\,  |\tr(\rho^{(V)}_\theta L^{(V)}_\theta P_j)|^2\\
\leq &\, \tr(L^{(V)}_\theta \rho^{(V)}_\theta L^{(V)}_\theta P_j) \tr(\rho^{(V)}_\theta P_j)\;.
\end{split}
\ee
Therefore,
\be\label{step1}
\begin{split}
\mathcal F_C^{(V)}(\theta) \leq &\, \sum_{j\in \mathcal J^*}  \tr(L^{(V)}_\theta \rho^{(V)}_\theta L^{(V)}_\theta P_j) 
\leq  \sum_{j\in \mathcal J}  \tr(L^{(V)}_\theta \rho^{(V)}_\theta L^{(V)}_\theta P_j) = \tr[\rho^{(V)}_\theta (L^{(V)}_\theta)^2]\;.
\end{split}
\ee
Taking the maximum over the initial preparation, 
\be\label{temp1}
\underset{\rho_0}{\text{max}}\;\mathcal F_C^{(V)}(\theta) \leq \underset{\rho_0}{\text{max}}\;  \tr[\rho^{(V)}_\theta (L^{(V)}_\theta)^2]\;.
\ee
By convexity, the maximum of the expression on the RHS is achieved when the system is prepared in a pure state. Let us set $\rho_0 = \ket{\psi_0}\bra{\psi_0}$. One can then rewrite it as
\be
\tr[\rho^{(V)}_\theta (L^{(V)}_\theta)^2] \big |_{\rho_0 = \ket{\psi_0}\bra{\psi_0}} = 4\,\text{Var}_{\ket{\psi_0}}(\tilde{U}^{(V)\dagger} \mathfrak g_\theta[\tilde{U}^{(V)}]\, \tilde{U}^{(V)})\;,
\ee
where 
\be\label{deco}
\mathfrak g_\theta[\tilde{U}^{(V)}] = \mathfrak g_{\theta}[S] + (S V)\,\mathfrak g_\theta[U_t]\,(S V)^\dagger
\ee
is the local generator of $\tilde{U}^{(V)}$. By Popoviciu's inequality we have,   
\be\label{s2in}
\underset{\rho_0}{\text{max}}\;\mathcal F_C^{(V)}(\theta) \leq [\sigma(\mathfrak g_{\theta}[S] + (S V)\,\mathfrak g_\theta[U_t]\,(S V)^\dagger)]^2\;,
\ee
and after maximizing over the unitary control $V$,
\be\label{rhs1}
\mathcal G(\theta) \leq \underset{V\in U(d)}{\text{max}} [\sigma(\mathfrak g_{\theta}[S] + (S V)\,\mathfrak g_\theta[U_t]\,(S V)^\dagger)]^2\;.
\ee
The above maximization may be carried out explicitly. To this aim we employ 
the following lemma: the maximum spectral gap of the sum of any two Hermitian matrices with given spectra is equal to the sum of their spectral gaps, i.e.
\be\label{lemmaBCR}
\underset{U_1,U_2\in U(d)}{\text{max}}\,\sigma(U_1 M_1 U_1^\dagger +U_2 M_2 U_2^\dagger) = \sigma(M_1) + \sigma(M_2)\;,\qquad M_1,M_2 \in \mathsf{Her}_d(\mathbb C)\;.
\ee
See Ref.~\cite{seveso_estimation_2017} for a proof. From Eq.~\eqref{lemmaBCR}, Eq.~\eqref{thmBCR1} follows immediately.  
\end{proof}
\end{thm}
\par
Let us now discuss tightness of inequality \eqref{thmBCR1}. The proof of Thm.~\ref{thm1BCR} can be broken down into three main steps:
\begin{itemize}
\item[\textbf{(S1)}] In Eq.~\eqref{step1}, the Fisher information $\mathcal F_C^{(V)}(\theta)$  was bounded from above. This step actually made use of three different inequalities: the inequality $\Re z \leq |z|$ (on the first line of Eq.~\eqref{step1ineq12}), the Cauchy-Schwarz inequality (on the second line of Eq.~\eqref{step1ineq12}) and the inequality on the second line of Eq.~\eqref{step1}, which follows from
\be\label{thirdIneq}
\sum_{j\in \mathcal J\setminus \mathcal J^*} \tr(L^{(V)}_\theta \rho^{(V)}_\theta L^{(V)}_\theta P_j) \geq 0\;.
\ee
\item[\textbf{(S2)}] Next, the quantity on the RHS of Eq.~\eqref{temp1} was maximized over the initial preparation $\rho_0$, which led to Eq.~\eqref{s2in}.
\item[\textbf{(S3)}] Finally, maximization over the unitary control $V$ was performed.  
\end{itemize}
Steps \textbf{(S2)} and \textbf{(S3)} are proper maximizations, that can be made tight by implementing the optimal control $V^{(opt)}$ and the optimal initial preparation $\ket{\psi_0^{(opt)}}$. It is easy to check that the optimal control has the form
\be\label{optctrl}
V^{(opt)} = S^\dagger R_1^\dagger R_2\;,
\ee 
where $R_1$ (\emph{resp.}, $R_2$) is the similarity transformation that diagonalizes $\mathfrak g_\theta[S]$ (\emph{resp.}, $\mathfrak g_\theta[U_t]$), with eigenvalues ordered decreasingly, i.e.
\be
\begin{split}
R_1 \mathfrak g_\theta[S] R_1^\dagger =&\, \text{diag}(\lambda_1(\mathfrak g_\theta[S]),\dots, \lambda_d(\mathfrak g_\theta[S]))\;,\\
R_2\,\! \mathfrak g_\theta[U_t] R_2^\dagger =&\, \text{diag}(\lambda_1(\mathfrak g_\theta[U_t]),\dots, \lambda_d(\mathfrak g_\theta[U_t]))\;.
\end{split}
\ee
Moreover, from Popoviciu's inequality, the optimal initial preparation is
\be\label{optPrepU}
\ket{\psi_0^{(opt)}} = \frac{1}{\sqrt 2}\, \tilde{U}^{(V^{(opt)})\dagger} [\ket{\lambda_1(\mathfrak g_\theta[\tilde{U}^{(V^{(opt)})}])} + e^{i\phi} \ket{\lambda_d(\mathfrak g_\theta[\tilde{U}^{(V^{(opt)})}])}]\;,\qquad \phi\in\mathbb R\;,
\ee
where $\tilde{U}^{(V^{(opt)})} = S V^{(opt)}U_t$. The previous expression for $\ket{\psi_0^{(opt)}}$ can be slightly simplified by noticing that the extremal eigenvalues of the generator of $\tilde{U}^{(V^{(opt)})}$ coincide with the extremal eigenvalues of the generator of $S$. This can be proven as follows. From Eq.~\eqref{deco} and Eq.~\eqref{optctrl}, the generator of $\tilde{U}^{(V^{(opt)})}$ can be written as
\be\label{genIdent}
\mathfrak g_\theta[\tilde{U}^{(V^{(opt)})}] = \mathfrak g_\theta[S] + R_1^\dagger R_2 \mathfrak g_\theta[U_t] R_2^\dagger R_1 = R_1^\dagger D R_1\;, 
\ee 
where $D$ is the diagonal matrix
\be
D = \text{diag}[\lambda_1(\mathfrak g_\theta[S])+\lambda_1(\mathfrak g_\theta[U_t]),\dots,\lambda_d(\mathfrak g_\theta[S])+\lambda_d(\mathfrak g_\theta[U_t])]\;.
\ee
Therefore, the extremal eigenvectors of $\mathfrak g_\theta[\tilde{U}^{(V^{(opt)})}]$ are given by 
\be
\ket{\lambda_1(\mathfrak g_\theta[\tilde{U}^{(V^{(opt)})}])} = R_1^\dagger \ket{0}\;,
\qquad\quad  \ket{\lambda_d(\mathfrak g_\theta[\tilde{U}^{(V^{(opt)})}])} = R_1^\dagger \ket{d-1}\;.
\ee 
But, by the very definition of $R_1$, $R_1^\dagger \ket{0} = \ket{\lambda_1(\mathfrak g_\theta[S])}$ and $R_1^\dagger \ket{d-1} = \ket{\lambda_d(\mathfrak g_\theta[S])}$, which establishes our claim. One may thus write
\be
\label{optprep}
\ket{\psi_0^{(opt)}} = \frac{1}{\sqrt 2}\, \tilde{U}^{(V^{(opt)})\dagger} [\ket{\lambda_1(\mathfrak g_\theta[S])} + e^{i\phi} \ket{\lambda_d(\mathfrak g_\theta[S])}]\;,\qquad \phi\in\mathbb R\;.
\ee
Proving tightness of inequality \eqref{thmBCR1} is therefore equivalent to proving that of step \textbf{(S1)}, under the constraints that the control and the initial preparation are chosen according to Eq.~\eqref{optctrl} and Eq.~\eqref{optprep}, respectively. Let us first consider the majorization based on the Cauchy-Schwarz inequality, which is saturated if and only if, $\forall j \in \mathcal J^*$, there exist complex numbers $\{\alpha_j\}$ such that
\be\label{condCSprop}
\sqrt{\rho_\theta^{(V^{(opt)})}} P_j = \alpha_j  \sqrt{\rho_\theta^{(V^{(opt)})}}L_\theta^{(V^{(opt)})}  P_j\;.
\ee
When the model is pure, condition \eqref{condCSprop} is automatically satisfied since it reduces to 
\be
\braket{\psi_\theta^{(V^{(opt)})}|j}\, \ket{ \psi_\theta^{(V^{(opt)})}}\bra{j}= \alpha_j \braket{\psi_\theta^{(V^{(opt)})}|L_\theta^{(V^{(opt)})}|j}\, \ket{ \psi_\theta^{(V^{(opt)})}}\bra{j}\;
\ee
(where we have set $\ket{\psi_\theta^{(V^{(opt)})}}\eqd \tilde{U}^{(V^{(opt)})}\ket{\psi_0^{(opt)}}$), which implies
\be
\alpha_j = \frac{\braket{\psi_\theta^{(V^{(opt)})}|L_\theta^{(V^{(opt)})}|j}}{\braket{\psi_\theta^{(V^{(opt)})}|j}}\;.
\ee
The remaining two inequalities used in step \textbf{(S1)} cannot be saturated without making further assumptions about the Hamiltonian $H_\theta$. For the inequality $\Re z \leq |z|$ to be tight, one should have, $\forall j \in \mathcal J^*$,
\be\label{relSat}
\Im \left[\braket{j|L_\theta^{(V^{(opt)})}|\psi_\theta^{(V^{(opt)})}} \braket{\psi_\theta^{(V^{(opt)})}|j}\right] = 0\;.
\ee
Upon writing explicitly the SLD $L_\theta^{(V^{(opt)})}$ and
using the optimal preparation given in Eq.~\eqref{optPrepU}, one may
prove that the inequality is tight provided that 
\be
|\braket{j|\lambda_1(\mathfrak g_\theta[S])}| = |\braket{j|\lambda_d(\mathfrak g_\theta[S])}|\;,\qquad \forall j \in \mathcal J^*\;, \label{pth}
\ee
i.e., the extremal eigenvectors of the generator of $S$, written in the computational basis, are such that corresponding entries have the same complex moduli. 
\par
It remains to discuss tightness of inequality \eqref{thirdIneq}. Let $j \in \mathcal J \setminus \mathcal J^*$. This is equivalent to $\braket{j|\psi_\theta^{(V^{(opt)})}}=0$ where, as before,
\be\label{secondeqn}
\ket{\psi_\theta^{(V^{(opt)})}} = \frac{1}{\sqrt 2}\,  [\ket{\lambda_1(\mathfrak g_\theta[\tilde{U}^{(V^{(opt)})}])} + e^{i\phi} \ket{\lambda_d(\mathfrak g_\theta[\tilde{U}^{(V^{(opt)})}])}]\;.
\ee 
For $\braket{j|\psi_\theta^{(V^{(opt)})}}=0$ to hold, there are two possibilities: either both
\be\label{onlyPoss}
\braket{j|\lambda_1(\mathfrak g_\theta[\tilde{U}^{(V^{(opt)})}])} = 0\quad \text{and} \quad \braket{j|\lambda_d(\mathfrak g_\theta[\tilde{U}^{(V^{(opt)})}])} = 0\;;
\ee
or they are different from zero, have the same moduli and the correct phase difference to cancel each other out. This last possibility can be excluded since the phase $\phi$ is arbitrary and can always be set such that no cancellation occurs. So the only possibility is for Eq.~\eqref{onlyPoss} to hold. Now, to prove tightness, one should show that
\be
\braket{j|L_\theta^{(V^{(opt)})}|\psi_\theta^{(V^{(opt)})}} = 0\;,\qquad\quad \forall j \in \mathcal J \setminus \mathcal J^*.
\ee
Using Eq.~\eqref{pth} and \eqref{secondeqn}, one arrives at the equivalent condition
\be
\lambda_1(\mathfrak g_\theta[\tilde{U}^{(V^{(opt)})}]) \braket{j|\lambda_1(\mathfrak g_\theta[\tilde{U}^{(V^{(opt)})}])} + \lambda_d(\mathfrak g_\theta[\tilde{U}^{(V^{(opt)})}])\, e^{i\phi} \braket{j|\lambda_d(\mathfrak g_\theta[\tilde{U}^{(V^{(opt)})}])}=0\;,
\ee
which is trivially satisfied because of Eq.~\eqref{onlyPoss}. Thus, no additional assumption is needed for equality to hold in Eq.~\eqref{thirdIneq}. We summarize our results via the following proposition.

\begin{prop}\label{usefulProp}
If $H_\theta$ is such that the extremal eigenvectors of the generator of its diagonalizing matrix $S$ satisfy the condition
\be\label{condBCR}
|\braket{j|\lambda_1(\mathfrak g_\theta[S])}| = |\braket{j|\lambda_d(\mathfrak g_\theta[S])}|\;,\qquad \forall j \in \mathcal J^*\;,
\ee
then the maximum Fisher information extractable via controlled energy measurements is
\be\label{resDefBCR}
\mathcal G(\theta) = \left[\sigma(\mathfrak g_\theta[U_t]) + \sigma(\mathfrak g_\theta[S])\right]^2\;.
\ee
The optimal preparation is given by Eq.~\eqref{optPrepU} and the optimal control by Eq.~\eqref{optctrl}.
\end{prop}
The condition imposed by Eq.~\eqref{condBCR} on $H_\theta$ may seem quite restricting. However, it turns out to be satisfied for many Hamiltonians of practical use in quantum metrology, see the examples discussed in Sect.~\eqref{examples}. Eq.~\eqref{resDefBCR} thus often provides a way to directly compute $\mathcal G(\theta)$, without the need of any optimization procedure.
\section{Metrological applications}\label{algo}
In this section, we discuss how to implement controlled energy measurements in a realistic metrological scenario. In principle, a controlled energy measurement requires to apply a unitary control $V$,
 and then to measure the energy projectively. The question is how to perform a projective measurement of the Hamiltonian when the Hamiltonian is not fully known. The problem has first been investigated in Refs.~\cite{nakayama_quantum_2015,matsuzaki_projective_2017}. In the following, we associate to each controlled energy measurement $\mathcal M_\theta^{(V)}$ a family of measurements, called \emph{realistic controlled energy measurements},  denoted by $\mathcal M_{n,m}^{(V)}$ (with $n,m \in\mathbb N$), that are experimentally feasible and allow to approximate $\mathcal M_\theta^{(V)}$ to any desired level of accuracy (in the sense that, as $n,m\to \infty$ the probability measure of $\mathcal M_{n,m}^{(V)}$ converges to that of $\mathcal M_\theta^{(V)}$). 
Our exposition can be divided into two parts. First, we describe a simplified version, denoted by ${\mathcal M}_{n}^{(V)}$, which is based on the \emph{phase estimation} algorithm \cite{temme_quantum_2011,riera_thermalization_2012,kitaev_quantum_1996}. It is assumed that the experimentalist can implement the controlled time-evolution operator 
\be
C_{U_t} \eqd \ket{0}\bra{0} \otimes \mathbb {I}_d + \ket{1}\bra{1} \otimes U_t\;.
\ee 
This is an unrealistic assumption, since $C_{U_t}$ still depends on the true value of the parameter via $U_t$. Next, we remove such assumption, which will lead to the introduction of realistic controlled energy measurements. 
\begin{figure}[h!]
\centering
\includegraphics[width=0.99\textwidth]{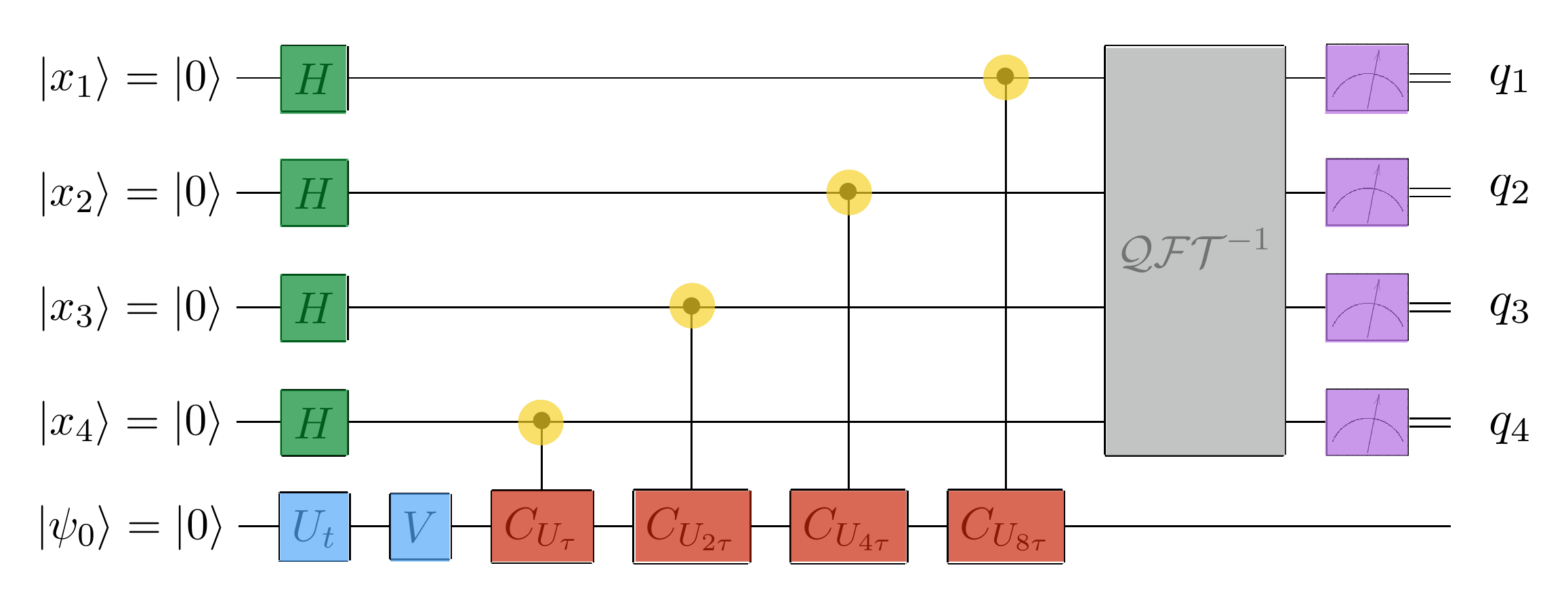}
\caption{Circuit diagram of ${\mathcal M}_{n}^{(V)}$ with $n=4$ control qubits. In a realistic implementation of controlled energy measurement, one 
replaces each box $\mathcal C_{U_\tau}$ by its effective implementation, i.e. by $m$ repeated applications of the transformation 
$\Gamma_{U_{\tau/m}}$,  defined in Eq.~\eqref{biggamma}.}
\end{figure}
In order to implement ${\mathcal M}_{n}^{(V)}$, one introduces $n$ control qubits, each one having Hilbert space $\mathcal H_c = \mathbb C^2$. The total Hilbert space is thus $\mathcal H_c^{\otimes n}\otimes \mathcal H$, with $\mathcal H=\mathbb C^d$ the Hilbert space of the original system. 
All the control qubits are initially prepared in their ground state
 $\ket{0}$, such that at time $t=0$ the state of the total system  is $\ket{0\mydots 0} \bra{0\mydots 0}\otimes \rho_0$. Then, 
 a Hadamard gate is applied to each control qubit, i.e. $\ket{0} \to H \ket{0} =(\ket{0} +\ket{1})/\sqrt 2$ and the parameter is encoded into the model $\rho_\theta = U_t \rho_0 U_t^\dagger$. Next, the unitary control $V$ is applied. At time $t$, the state of the system is thus given by
\be
\frac{1}{2^{n}}\,\sum_{x,y \in \{0,1\}^{\times n}}
\ket{x_1\mydots\, x_n} \bra{y_1\mydots\, y_n} \otimes V \rho_\theta V^\dagger\;,
\ee
where $x$ stands for the generic binary $n$-string $x_1\mydots\, x_n$ and $y$ for the binary string $y_1\mydots\, y_n$. 
\par
Next, given an arbitrary unitary $U$ on $\mathcal H$, we define the superoperator $\mathcal C_U$ as follows, 
\be\label{CU}
\mathcal C_U [\rho] \eqd C_U \rho\, C_U^\dagger\;. 
\ee
For $l=1,\mydots,n$, the $n$ superoperators $\mathcal C_{U_{\tau}^{2^{l-1}}}$ 
couple  the $l^{\text{th}}$ control qubit to the main system (here $\tau$ represents a free parameter, which corresponds to the typical time of the measurement process). In particular, when $\mathcal C_{U_{\tau}^{2^{l-1}}}$ is applied to $\rho_l \eqd \ket{x_l}\bra{y_l}\otimes V \rho_\theta V^\dagger$, one obtains
\be
\mathcal C_{U_{\tau}^{2^{l-1}}}\left[\rho_l \right] =  \ket{x_l} \bra{y_l} \otimes  U_\tau^{x_l 2^{l-1}} V \rho_\theta V^\dagger \left(U^\dagger_\tau\right)^{y_l 2^{l-1}}\;.
\ee
Denoting by $X = x_1 + 2\cdot x_2 +\mydots + 2^{n-1}\cdot x_n $ the decimal representation of the binary string $x$, one obtains
\be\label{ts1}
\frac{1}{2^{n}}\, \sum_{X=0}^{2^n-1}\, \sum_{Y=0}^{2^n-1} \ket{x} \bra{y} \otimes U_\tau^X V \rho_\theta V^\dagger (U^\dagger_\tau)^Y\;.
\ee
Let us now expand $V\rho_\theta V^\dagger$ on the energy eigenbasis, i.e.
\be
V\rho_\theta V^\dagger = \sum_{j=0}^{d-1} \sum_{k=0}^{d-1} c_{jk} \ket{\xi_{j,\,\theta}} \bra{\xi_{k,\,\theta}}\;.
\ee
Eq.~\eqref{ts1} then becomes
\be
\frac{1}{2^{n}}\! \sum_{j,k=0}^{d-1}\, \sum_{X,Y=0}^{2^n-1} c_{jk}\,e^{-i\tau (X \xi_{j,\,\theta}-Y \xi_{k,\,\theta})}
\ket{x}\! \bra{y} \otimes \ket{\xi_{j,\,\theta}}\! \bra{\xi_{k,\,\theta}}\;.
\ee
The subsequent step of the protocol involves the use of inverse 
quantum Fourier transform $\mathcal{QFT}^{-1}$ on the set of 
$n$ control qubits. The action of $\mathcal{QFT}^{-1}$ on 
the computational basis (of $\mathcal H_c^{\otimes n}$) is given 
by:
\be
\mathcal{QFT}^{-1}\ket{x} = \frac{1}{2^{n/2}}\, \sum_{Q=0}^{2^{n}-1} e^{-\frac{2\pi i X Q}{2^n}}\ket{q}\,.
\ee
and thus the total state of the system after $\mathcal{QFT}^{-1}$ may be written as
\be
\frac{1}{2^{2n}}\, \sum_{j,k=0}^{d-1}\,\sum_{X,Y=0}^{2^n-1}\,\sum_{Q,P=0}^{2^n-1}\, \tilde c_{jk}\, \ket{q}\bra{p} \otimes \ket{\xi_{j,\,\theta}}\bra{\xi_{k,\,\theta}}\;.
\ee
where 
\be
\tilde c_{jk} = c_{jk}\, e^{-iX\left(\tau \xi_{j,\,\theta}+\frac{2\pi Q}{2^n}\right)}\, e^{iY\left(\tau \xi_{k,\,\theta}+\frac{2\pi P}{2^n}\right)}\;.
\ee
The final step consists in a read-out, i.e. one performs a 
measurement (in the computational basis) on the $n$ control qubits . 
The probability $\text{Pr}_{\theta}(q)$ of obtaining the (binary) 
string $q$ as outcome is given by 
\be\label{pp2}
\text{Pr}_{\theta}(q) = \frac{1}{2^{2n}} \sum_{j=0}^{d-1} \, \sum_{X,Y=0}^{2^n-1} \text{Pr}_\theta(\xi_{j,\theta})\, e^{-i(X-Y)\alpha_{j,Q}}\;,
\ee
where
\be
\alpha_{j,Q} \eqd \tau \xi_{j,\,\theta} + \frac{2\pi Q}{2^n}\;,\qquad \text{Pr}_\theta(\xi_{j,\theta}) = \braket{\xi_{j,\,\theta}|V \rho_\theta V^\dagger|\xi_{j,\,\theta}} \;.
\ee
After straightforward manipulation, Eq.~\eqref{pp2} can also be written as
\be\label{pqsin}
\text{Pr}_{\theta}(q) = \sum_{j=0}^{d-1} \text{Pr}_\theta(\xi_{j,\theta}) \left(\frac{1}{2^n}\frac{\sin(2^n \alpha_{j,Q}/2)}{\sin(\alpha_{j,Q}/2)}\right)^2\;.
\ee
In the limit $n\to \infty$, $\text{Pr}_{\theta}(q)$ converges to the probability $\text{Pr}_{\theta}(\xi_{j,\theta})$, which corresponds to a controlled energy measurement $\mathcal M_{\theta}^{(V)}$. 
\begin{figure}[h]
\centering
\includegraphics[width=0.99\columnwidth]{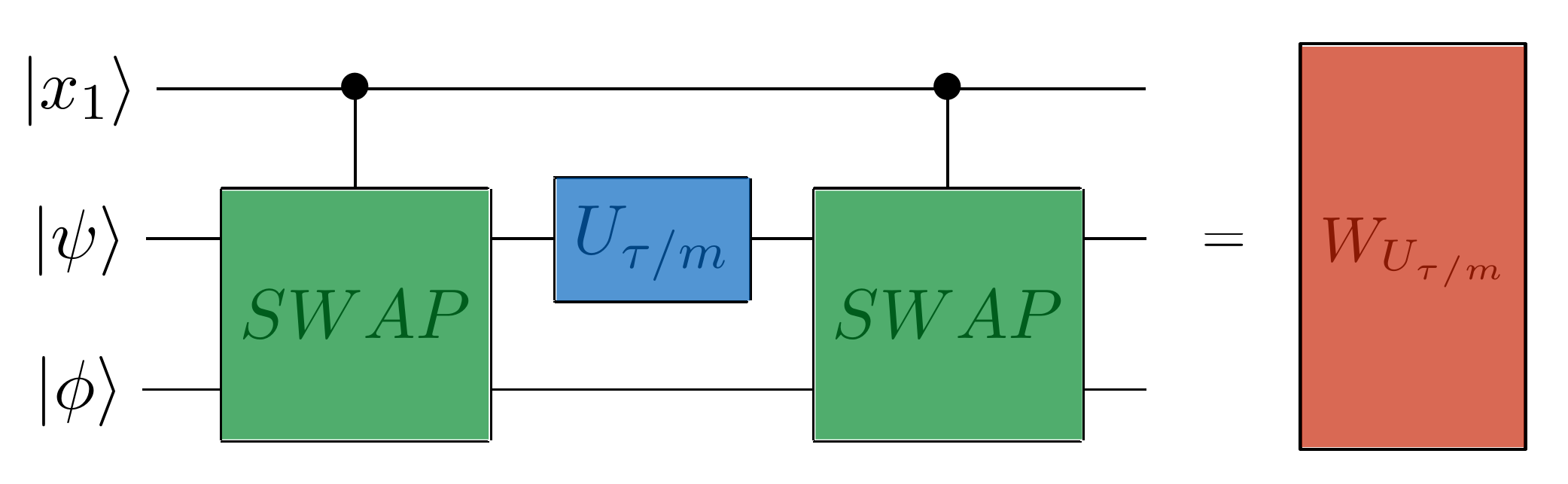}
\caption{Circuit diagram of $W_{U_{\tau/m}}$. \label{w}}
\end{figure}\par
In order to obtain acontrolled energy measurement $\mathcal M_{n,m}^{(V^)}$ 
in a realistic scenario, one exploits ${\mathcal M}_{n}^{(V)}$ and 
implements the controlled time-evolution operator $C_{U_t}$ using a 
quantum subroutine referred to as \emph{universal controllization} \cite{nakayama_quantum_2015}. In order to briefly illustrate the
protocol let let us address the case $l=1$ and consider the problem to approximate the action of $\mathcal C_{U_\tau}$ on the state $\rho_1 = \ket{x_1}\bra{y_1} \otimes V \rho_\theta V^\dagger$. The transformation 
$\mathcal C_{U_\tau}$ is obtained (i.e. replaced) by $m$ applications of the superoperator $\Gamma_{U_{\tau/m}}$, constructed as follows. 
At first, an ancilla system with the same dimensionality as the main system, is introduced. The total Hilbert space is $\mathcal H_c^{\otimes n} \otimes \mathcal H \otimes \mathcal H_a$, with $\mathcal H_a=\mathbb C^d$. The ancillary system is then prepared in a maximally mixed state: The state of 
the first control qubit, the main system and the ancilla (before application of $\mathcal C_{U_\tau}$) is thus given bu $\rho'_1 = \ket{x_1}\bra{y_1} \otimes V \rho_\theta V^\dagger \otimes \mathbb I_d/d$. Let us now consider the 
quantum operation,
\be
W_{U_\tau} \eqd C_{SWAP} (\mathbb I_2 \otimes U_\tau \otimes \mathbb I_d)\, C_{SWAP}\;,
\ee 
where $C_{SWAP}$ is the controlled-SWAP gate acting  on $\mathcal H_c \otimes \mathcal H \otimes \mathcal H_a$ as
\be
C_{SWAP}(\ket 0 \otimes \ket \psi \otimes \ket \phi) = \ket 0 \otimes \ket \phi \otimes \ket \psi\;,\qquad C_{SWAP}(\ket 1 \otimes \ket \psi \otimes \ket \phi) = \ket 0 \otimes \ket \psi \otimes \ket \phi\;. 
\ee
At this point, it is crucial to remark that for 
the realization (implementation) of the transformation $W_{U_\tau}$ 
we do not need to know the form the Hamiltonian, since only 
the \emph{uncontrolled} version of the time-evolution operator $U_\tau$ is required. Let us  divide $\tau$ into $m$ subintervals of duration 
$\tau/m$. During each subinterval, $W_{U_{\tau/m}}$ is applied; then the ancilla is traced out and finally it is reset to its initial state. As for example: after the first interval, one obtains $\Gamma_{U_{\tau/m}}[\rho_1]\otimes \mathbb I_d/d$, where 
\be\label{biggamma}
\Gamma_{U_{\tau/m}}[\rho_1] \eqd \tr_{\mathcal H_a}\left(W_{U_{\tau/m}} \rho_1'\,  W^\dagger_{U_{\tau/m}}\right)\;.
\ee
A simple computation reveals that
\be
\Gamma_{U_{\tau/m}}[\rho_1] = \frac{1}{d}\tr\left(U_{\tau/m}^{y_1-x_1}\right)\, \mathcal C_{U_{\tau/m}}[\rho_1]\;. 
\ee
For future convenience, we write
\be
\frac{1}{d}\,\tr\left(U_{\tau/m}\right) = a_{\tau/m}\, e^{i\phi_{\tau/m}}\;,
\ee
where $a_{\tau/m}\in \mathbb R^+$ and $\phi_{\tau/m}\in \mathbb R$. Note that, since $x_1-y_1 \in \{-1,0,1\}$, one can write
\be
\Gamma_{U_{\tau/m}}^m[\rho_1] = a_{\tau/m}^{|x_1-y_1|m}\, e^{i(y_1-x_1)m\phi_{\tau/m}}\,\mathcal C_{U_{\tau}}[\rho_1]\;.
\ee
Universal controllization thus replaces $\mathcal C_{U_{\tau}}$ with $\Gamma^m_{U_{\tau/m}}$. In the limit $m\to\infty$, it can be proven that the error
\be
\epsilon_m \eqd \left[\tr\left(U_{\tau/m}\right)/d\right]^m-1
\ee
tends to zero. A realistic controlled energy measurement is  obtained by substituting each application of $\mathcal C_{U^{2^{l-1}}_{\tau}}$ by $2^{l-1}m$ applications of  $\Gamma_{U_{\tau/m}}$. For instance, instead of Eq.~\eqref{ts1}, one would have
\be
\frac{1}{2^n} \sum_{X,Y=0}^{2^n-1} \pi_{X,Y} e^{i(Y-X)m \phi_{\tau/m}} \ket{x}\!\bra{y} \otimes U_\tau^X V \rho_\theta V^\dagger (U^\dagger_\tau)^Y\,,
\ee 
where 
\be
\pi_{X,Y}\eqd \prod_{l=1}^n a_{\tau/m}^{|x_l-y_l| 2^{l-1} m }\;.   
\ee
After applying the inverse quantum Fourier transform and measuring in the computational basis, the probability of obtaining the outcome $q\in \{0,1\}^{\times n}$ is
\be\label{pqfin}
\text{Pr}_\theta(q) = \frac{1}{2^{2n}} \sum_{j=0}^{d-1} \text{Pr}_\theta(\xi_{j,\theta}) \sum_{X,Y=0}^{2^n-1} \pi_{X,Y}\, e^{i (Y-X) \beta_{j,Q}}\;, 
\ee
with 
\be
\beta_{j,Q} \eqd \alpha_{j,Q} + m \phi_{\tau/m}\;.
\ee
Eq.~\eqref{pqfin} can be further expanded by rewriting it as follows, 
\be\label{pqcos}
\begin{split}
\text{Pr}_\theta(q) = \frac{1}{2^{2n}} \sum_{j=0}^{d-1} \text{Pr}_\theta(\xi_{j,\theta}) \prod_{l=1}^n \sum_{u,v=0}^1  a_{\tau/m}^{|u-v| 2^{l-1} m} e^{i(v-u) 2^{l-1}\beta_{j,Q}}\\
= \frac{1}{2^n} \sum_{j=0}^{d-1} \text{Pr}_\theta(\xi_{j,\theta})  \prod_{l=1}^n \left[1+ a_{\tau/m}^{ 2^{l-1} m}\cos\left(2^{l-1} \beta_{j,Q}\right) \right]\;.
\end{split}
\ee
If $m\to \infty$, then $\phi_{\tau/m}\to 0$ and $a_{\tau/m}\to 1$, so that Eq.~\eqref{pqcos} converges to Eq.~\eqref{pqsin}. In conclusion, a realistic controlled energy measurement allows to approximate to any desired precision a controlled energy measurement $\mathcal M^{(V)}_{\theta}$, without requiring any a priori knowledge about the parameter $\theta$. 
\section{Examples}\label{examples}
In this section, we work out a collection of examples. For each example, we compute the QFI $\mathcal F_Q(\theta)$ and compare it with  $\mathcal G(\theta)$. We will find that, in general, $\mathcal G(\theta)$ majorizes $\mathcal F_Q(\theta)$, thus controlled energy measurements lead to a precision enhancement. Moreover, we study numerically the performance of realistic controlled energy measurements $\mathcal M_{n,m}^{(V^{(opt)})}$. From the previous section, as $n, m \to \infty$, $\mathcal M_{n,m}^{(V^{(opt)})}$ converges to  $\mathcal M_\theta^{(V^{(opt)})}$, and thus its Fisher information also converges to $\mathcal G_\theta$. We will show that, already for relatively small values of $n$ and $m$, realistic controlled energy measurements perform very close to the the ultimate bound $\mathcal G_\theta$. 
\subsection{Estimation of the direction of a magnetic field}\label{s:theta}
\begin{figure}[h!]
\centering
\includegraphics[width=0.99\columnwidth]{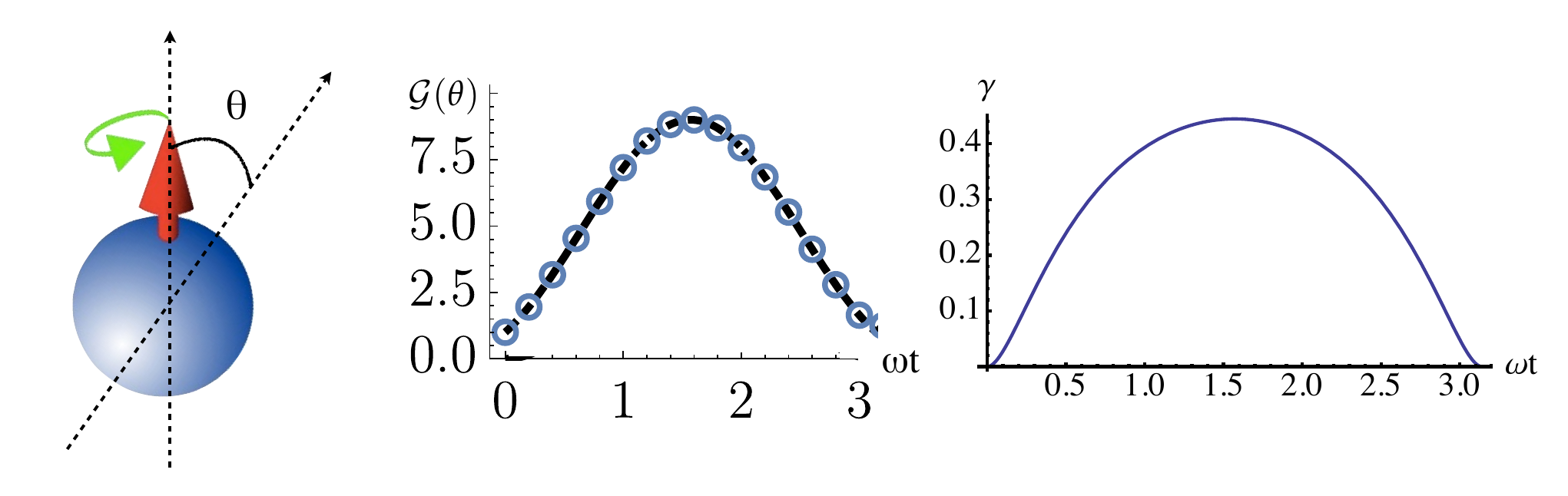}
\caption{Left: schematic diagrams of the qubit-based estimation 
of the direction of a magnetic field.  Center:  the line corresponds to $\mathcal G(\theta)$, computed by Eq.~\eqref{gximagf}, whereas the circular marks correspond to values of $\mathcal G(\theta)$ computed by the numerical optimization involved in its definition, see \eqref{cefid}. Right: comparison between the optimal Braunstein-Caves measurement and the optimal controlled energy measurement, the plot shows the ratio between  the QFI, optimized over the initial preparation and 
$\mathcal G(\theta)$, computed by Eq.~\eqref{gximagf}.\label{cmp1}}
\end{figure}

Let us consider a situation where the parameter of interest is direction of a magnetic field. More precisely, we want to estimate that
the polar angular direction $\theta$ of an external magnetic field, whose 
magnitude $B$ is known. The probing system is a two-level atom, with Hilbert space $\mathcal H = \mathbb C^2$ and Hamiltonian is  $H_\theta=\omega(\cos\theta\, \sigma_z + \sin\theta\, \sigma_x)$. The energy splitting 
$\omega$ is proportional to the magnitude $B$ of the field and it is 
thus known. At time $t=0$, the atom is initialized in its ground state: $\ket{\psi_0} = \ket 0$. At the generic time $t$, the state of the probe is $\ket{\psi_\theta}= U_t \ket{\psi_0}$, with $U_t \eqd \text{\emph{exp}}(-i H_\theta t)$, see the left panel of  Fig.~\ref{cmp1} for a schematic diagram.
\par
If an experimentalist is constrained to perform regular measurements, the best performance she can achieve is quantified by the QFI:
\be\label{qfimag}
\mathcal F_Q(\theta) = 4 \sin^2(\omega t) - \sin^2(2 \omega t) \sin^2\theta\;. 
\ee
Optimizing also over the initial preparation, 
 \be
 \underset{\ket{\psi_0}}{\text{\emph{max}}}\; \mathcal F_Q(\theta) = 4 \sin^2 (\omega t)\;.
 \ee
If instead the experimentalist is allowed to implement only controlled energy
measurements, the maximum Fisher information that she can extract is given 
by $\mathcal G(\theta)$. To compute  $\mathcal G(\theta)$, one first computes 
the matrix $S$, built from the eigenvectors of $H_\theta$, and its generator 
$\mathfrak g_{\theta}[S]$
\be
S = \left(\begin{matrix}
-sc(\theta) \sin \frac{\theta}{2} & sc(\theta) \cos\frac{\theta}{2}\\
ss(\theta) \cos\frac{\theta}{2}  &  ss(\theta) \sin\frac{\theta}{2}
\end{matrix}\right)\, \qquad \mathfrak g_{\theta}[S] = \left(\begin{matrix}
0&-\frac{i}{2}\,ss(\theta)\\
\frac{i}{2}\,ss(\theta)&0\\
\end{matrix}\right)\;.
\ee
where $sc(\theta)=\text{sgn}\left[\cos\frac{\theta}{2}\right]$,  $ss(\theta)=\text{sgn}\left[\sin\frac{\theta}{2}\right]$, and $\text{sgn}(x) \eqd |x|/x$. 
The extremal eigenvectors of $\mathfrak g_{\theta}[S]$ are then given by
\be
\ket{\lambda_1(\mathfrak g_{\theta}[S])}=\frac{1}{\sqrt 2}\;(-i,\;1)^t\;,\qquad \ket{\lambda_2(\mathfrak g_{\theta}[S])}=\frac{1}{\sqrt 2}\;(i,\;1)^t\;.
\ee
Since condition \eqref{condBCR} is satisfied, $\mathcal G(\theta)$ can be obtained via Prop.~\eqref{usefulProp}.
The explicit expressions for $U_t$ and its generator are 
\begin{align}
U_t = \left(\begin{matrix}
 A &  B\\
B  &  A^*\\
\end{matrix}\right)\,, \qquad \qquad 
\mathfrak g_\theta[U_t] & =
\left(\begin{matrix}
- C & D \\
D^* & C
\end{matrix}\right)\;, 
\end{align}
where
\bes
\begin{split}
A & = \cos\omega t- i\cos\theta\sin\omega t\;,\qquad\quad
C  = \frac{1}{2}\sin\theta\sin 2\omega\;, \\ 
B & = -i\sin\theta\sin\omega t  \;,\qquad\qquad\qquad
D  = \left(\cos\theta\cos \omega t-i\sin \omega t\right)\sin\omega t
\;.
\end{split}
\ees
One thus obtains
\be\label{gximagf}
\mathcal G(\theta) =  \Big(\,2|\sin(\omega t)| +1\,\Big)^2\;.
\ee
As an overall check, in the central panel of 
Fig.~\ref{cmp1} we report $\mathcal G(\theta)$ 
computed by Eq.~\eqref{gximagf} together with its values 
computed by numerical optimization from its definition \eqref{cefid}.
In the right panel we instead show a comparison of 
$\mathcal G(\theta)$ with the QFI in terms of the ratio 
$\gamma=\text{\emph{max}}_{\ket{\psi_0}} \mathcal F_Q(\theta)/\mathcal G(\theta)$, which is apparently below unit at all times.
\par
In order to check whether the above protocol may be of practical interest one
may also study numerically the performance of $\mathcal M_{n,m}^{(V^{(opt)})}$ (with $V^{(opt)}$ the optimal control of Eq.~\eqref{optctrl}). Recall that $n$ is the number of ancillary qubits needed to implement the phase estimation algorithm, while $m$ is the number of subintervals the  timescale $\tau$ is subdivided into. During each subinterval, the action of the controlled time-evolution operator $\mathcal C_{U_\tau}$ is approximated by applying $m$ times the superoperator $\Gamma_{U_{\tau/m}}$ of Eq.~\eqref{biggamma}. As $n,m\to \infty$, the probability measure of $\mathcal M_{n,m}^{(V^{(opt)})}$ converges to that of the optimal controlled energy measurement. Our results show that already for reasonably small values of the two parameters, say ~$n=6$, $m=3$, one is  close to the ultimate bound $\mathcal G(\theta)$.
\subsection{Estimation of a component of a magnetic field}\label{s:bx}
\begin{figure}[h!]
\centering
\includegraphics[width=0.99\columnwidth]{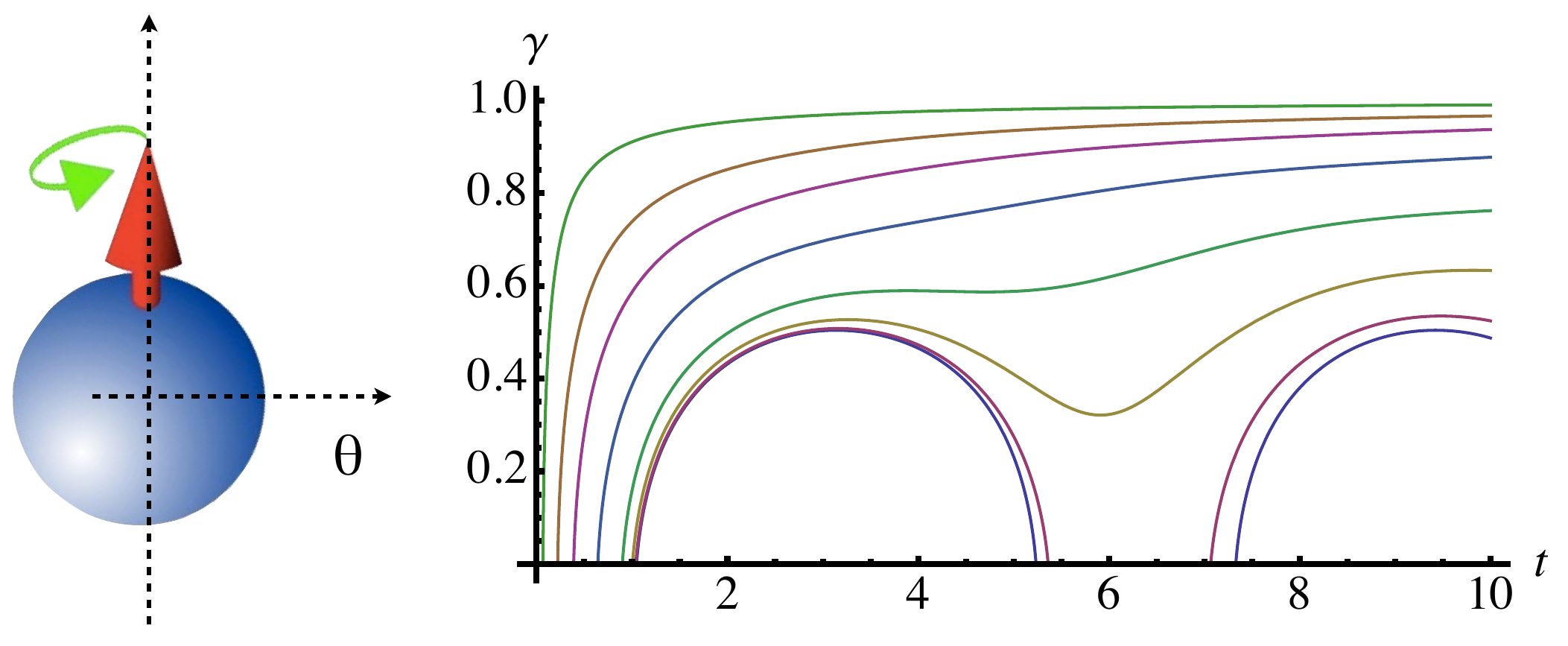}
\caption{Left: schematic diagrams of the qubit-based estimation 
the x-component of a magnetic field. Right:
the ratio $\gamma=\emph{max}_{\ket{\psi_0}}\mathcal F_Q(\theta)/\mathcal G(\theta)$ as a function of time for $\omega=1$ and different values of $\theta$. From lower to upper curves, we have $\theta=0, \pi/50,\pi/20, \pi/10, \pi/5, \pi/3, \pi/2$. Notice the loss of periodicity for higher values of $\theta$.
\label{f4_bx}}
\end{figure}

The parameter to be estimated is the component of a magnetic 
field along the $x$ direction. The probing system is again a two-level atom. The Hamiltonian is $H_\theta = -\omega \sigma_z + \theta \sigma_x$, with eigenvalues $\pm \Omega_\theta$ and $\Omega_\theta \coloneqq \sqrt{\omega^2+\theta^2}$. 
We report the matrices $U_t$ and $S$, with their corresponding generators. 
For $U_t$ and $\mathfrak g_\theta[U_t]$, one obtains
\begin{align}
U_t = \left(\begin{matrix}
 A &  B\\
B  &  A^*\\
\end{matrix}\right)\,, \qquad\quad 
\mathfrak g_\theta[U_t] & =
\left(\begin{matrix}
- C & D \\
D^* & C
\end{matrix}\right)\;, 
\end{align}
where
\bes
\begin{split}
A & =  \cos (\Omega_\theta t)+\frac{i \omega  
\sin (\Omega_\theta t)}{\Omega_\theta}\,, \quad
C  =  - \frac{ \omega\theta \,  [\sin (2\,\Omega_\theta t)-2\,\Omega_\theta t]}{2\,\Omega_\theta^3}\;,  \\
B & = -\frac{i \theta  \sin (\Omega_\theta t)}{\Omega_\theta}\,, \quad
D  = \frac{\sin (2\,\Omega_\theta t) \omega ^2-i \Omega_\theta \cos (2\,\Omega_\theta t) \omega +\Omega_\theta \left(2  t\theta^2+i \omega \right)}{2\,\Omega_\theta^3} 
\;.
\end{split}
\ees
For the matrix $S$ and its generator,
\be
\begingroup
\renewcommand*{\arraystretch}{1}
S  =\frac{1}{\sqrt{2\,\Omega_\theta}}
\left(\begin{matrix}
 -\frac{\omega + \Omega_\theta}{\sqrt{\Omega_\theta+\omega}}&\frac{\theta}{\sqrt{\Omega_\theta+\omega}}\\
 \frac{\theta}{\sqrt{\Omega_\theta+\omega}}&\frac{\theta}{\sqrt{\Omega_\theta-\omega}}
\end{matrix}\right)\;, \qquad \qquad 
\mathfrak g_{\theta}[S]  = \frac{i \omega }{2 \theta ^2}
\left(\begin{matrix}
 0 & 1 \\
 -1 & 0 \\
\end{matrix}\right)\;.
\endgroup
\ee
The maximum QFI is
\be\label{r1}
\underset{\ket{\psi_0}}{\text{\emph{max}}}\; \mathcal F_Q(\theta) =\frac{2}{\Omega_\theta^4} \Big[2\, \Omega_\theta^2\, t^2\theta^2-\omega ^2 \cos (2\,\Omega_\theta t)+\omega ^2\Big]\;.
\ee
Since the eigenvectors of $\mathfrak g_{\theta}[S]$ satisfy condition \eqref{condBCR}, $\mathcal G(\theta)$ can be computed directly and is 
given by 
\be\label{r2}
 \mathcal G(\theta)=\Bigg(\frac{\omega}{\Omega_\theta^2} + \frac{\sqrt{2 \left[2\,\Omega_\theta^2\, t^2\theta^2-\omega ^2 \cos (2\,\Omega_\theta t)+\omega ^2\right]}}{\Omega_\theta^2}\, \Bigg)^2\;,
\ee 
which is larger than the QFI at any time. In particular, we may write
\be
 \mathcal G(\theta) = \left(\frac{\omega}{\Omega_\theta^2} + \sqrt{\underset{\ket{\psi_0}}{\text{\emph{max}}}\; \mathcal F_Q(\theta)}\,\right)^2\,.
\ee
The ratio $\gamma=\text{\emph{max}}_{\ket{\psi_0}} \mathcal F_Q(\theta)/\mathcal G(\theta)$ may be written as $\gamma = 1- \omega/(\theta^2 t)$ for $\omega\ll 1$, whereas the difference between the difference between 
$\mathcal F_Q(\theta)$ and $\mathcal G(\theta)$ may be more 
pronounced in other regimes.  For $\omega\gg 1$, the ratio $\gamma$ 
oscillates at small times, and then it approaches unity for large times.
In Fig. \ref{f4_bx}, we show the ratio $\gamma$ as a 
function of time for $\omega=1$ and different values of 
$\theta$).
\subsection{Estimation of a weak magnetic field by spin-1 probes}
NV-center in diamond has been suggested as quantum probes to precisely 
estimate the magnitude of a weak magnetic field. The probing system is made 
of a nitrogen atom (N) inside a diamond crystal lattice, having a vacancy (V) in one of its neighboring sites. Two different classes of the defects 
are known and employed: the neutral state, usually referred to as $NV_0$, 
and the negatively-charged state $NV_{-}$. The second class $NV_-$ is the one exploited  in metrological applications, since it provides a spin triplet 
state which can be accurately prepared, manipulated with long coherence 
time, and finally read out by purely optical means \cite{rondin_magnetometry_2014}. Upon assuming that the interactions 
with the surrounding nuclear spins may be neglected, the 
Hamiltonian $H_{NV}\,$ governing the evolution of 
the triplet state is given by
\be
H_{NV} =  \mu\,\boldsymbol B \cdot \boldsymbol S + D\, S_z^2 + E\, (S_x^2-S_y^2)\;,
\ee
where the external magnetic field is denoted by $\boldsymbol B$ and $\boldsymbol S =(S_x, S_y, S_z)$ is a vector whose elements are the three spin 1 matrices:
\bes
\label{pms1}
S_x =\sqrt 2\, \left(\begin{matrix}
0&1&0\\
1&0&1\\
0&1&0\\
\end{matrix}\right)\;,  \qquad S_y = \sqrt 2 i\,
\left(\begin{matrix}
0&-1&0\\
1&0&-1\\
0&1&0\\
\end{matrix}\right)\;,\qquad 
 S_z  = 2 
\left(\begin{matrix}
1&0&0\\
0&0&0\\
0&0&-1\\
\end{matrix}\right)\;.
\ees
In the above formulas $\mu$ is the Bohr magneton and the couplings $D$ 
and $E$ are given by $D\apeq \pi \,\times\, 1.44\,\si{GHz}$ and 
$E\apeq \pi \,\times\, 50\, \si{kHz}$, respectively. Upon assuming that 
the magnetic field is weak, the two transverse components $B_x$ and $B_y$ may be neglected in comparison to the component $B_z$, which is aligned along the NV-center defect axis. By renaming $B_z$ as $\theta$, the Hamiltonian becomes
\be
H_\theta = \mu \theta S_z + D\, S_z^2 + E\, (S_x^2-S_y^2)\;.
\ee
The maximum QFI is
\be
 \underset{\ket{\psi_0}}{\text{\emph{max}}}\;\mathcal F_Q(\theta)=\frac{8  \mu ^2 \left[2\theta^2 \mu^2 t^2 \chi^2+E^2-E^2 \cos \left(4 \chi t\right)\right]}{\chi^4}\;,
\ee
where $\chi\eqd \sqrt{\theta^2\mu^2 +4 E^2}$. Instead, $\mathcal G(\theta)$ is given by
\begin{align}
\mathcal G(\theta)  & =  \left(\frac{2 E \mu }{\chi^2}+2\sqrt 2\mu\frac{ \sqrt{2\theta^2 \mu^2 t^2 \chi^2+E^2-E^2 \cos \left(4 \chi t\right)}}{\chi^2} \right)^2 \notag \\
  & =  \left(\frac{2 E \mu }{\chi^2}+ \sqrt{\underset{\ket{\psi_0}}{\text{\emph{max}}}\; \mathcal F_Q(\theta)}\right)^2 \geq \underset{\ket{\psi_0}}{\text{\emph{max}}}\; \mathcal F_Q(\theta)\,,
\end{align}
with the maximised QFI approaching the value of ${\cal G} (\theta)$ 
only in the limit $\theta \gg 1$.
\section{Conclusions}\label{s:out}
In this paper, we have addressed non-regular measurements as a novel 
resource for quantum metrology. In particular, we have analysed the 
family of controlled energy measurements and applied them to Hamiltonian parameter estimation problems. A controlled energy measurement is 
obtained by applying a unitary control and then performing a projective energy measurement. It is non-regular whenever the Hamiltonian depends 
non-linearly on the parameter $\theta$. 
\par
We have then maximized the Fisher information over the set of controlled energy measurements and initial preparations. The maximum, denoted by $\mathcal G(\theta)$, can be computed by the closed-form expression given in Eq.~\eqref{resDefBCR}, and it may be larger than the QFI of the corresponding 
regular statistical model. We have discussed how controlled energy 
measurements can be implemented in realistic scenarios, via an 
adaptation of the quantum phase estimation algorithm. 
\par
Finally, in order to to clarify the details of our 
estimation techniques, we have worked out a collection of examples, 
showing that a precision enhancement, compared with regular measurements, 
is often possible. In particular, we have emphasized that, if the parameter 
is not a simple phase, the quantum Fisher information no longer necessarily embodies the ultimate  precision limit. Our results show that precision of quantum metrological protocols is not necessarily bounded by the inverse of the quantum Fisher information, i.e. quantum enhanced estimation 
may be more precise than previously thought. We foresee further applications in the field of quantum sensing \cite{np17} and quantum probing \cite{qprob1,qprob2,qprob3,qprob4}
\section*{Acknowledgments}{The authors thanks Matteo Rossi, Francesco Albarelli, Marco Genoni, Claudia Benedetti, Stefano Olivares, Dario Tamascelli, C. M. Chandrasekar, Ilaria Pizio, Shivani Singh and Sholeh Razavian for interesting discussions. This work has been supported by SERB through project VJR/2017/000011. MGAP is a member of~GNFM-INdAM.}
\section*{Appendix: abbreviations and symbols used in this paper}
$ $ \par\noindent 
\begin{tabular}{@{}ll}
MSE & mean-square error \\
POVM & Positive operator-valued measure\\
FI & Fisher Information \\
SLD & Symmetric logarithmic derivative\\
QFI & Quantum Fisher Information
\end{tabular}
\begin{longtable}{p{1in} p{5in} }
  $\mathbb N$         & Set of positive integers \\
  $\mathbb N_0$       & Set of nonnegative integers \\
  $\mathbb R$         & Set of real numbers \\
  $\overline{\mathbb R}$   & Extended set of real numbers \\
  $\mathbb R_+$       & Set of nonnegative real numbers \\
  $\mathbb C$         & Set of complex numbers \\
  $|S|$               & Cardinality of a set \\
  $\mathscr P(S)$     & Power set of $S$  \\
  $\text{conv}(S)$    & Convex hull of a set of points $S$ \\
  $\text{vert}(\Pi)$  & Set of vertices of a convex polytope $\Pi$ \\
  $M_{ij}$    &   Element $ij$ of $M$ \\
  $M^t$    &   Transpose of a matrix $M$ \\
  $\text{spec($M$)}$    &   Spectrum of a matrix $M$ \\
  $\text{rk($M$)}$    &   Rank of a matrix $M$ \\
  $\sigma(M)$    &   Spectral gap of a matrix $M$ \\
  $\text{col($M$)}$    &   Set made up of the columns of a matrix $M$ \\
  $\text{diag}(\{\lambda_i\}_{i=1}^n)$    &   Diagonal matrix, with diagonal elements $\{\lambda_i\}_{i=1}^n$ \\
  $\text{im}_S(M)$    &   Image of a matrix $M$ on a set $S$\\
  $\mathcal M_{n,m}(\mathbb K)$     &   Set of $n\times m$ matrices over a field $\mathbb K$ \\
  $\mathsf{Her}_{n}(\mathbb K)$    &    Set of $n\times n$ Hermitian matrices  over a field $\mathbb K$ \\
  $\mathsf{Her}_{n}^+(\mathbb K)$   &   Set of $n\times n$ positive semi-definite Hermitian matrices over a field $\mathbb K$ \\
  $\mathbb I_n$   &   $n\times n$ identity matrix \\
  $\mathbb 0_n$   &   $n\times n$ zero matrix \\
  $\mathbb J_{p\times q}$   &   $p\times q$ matrix made up of all ones \\
  $\mathbb X,\,\mathbb Y\dots$    &   Classical random variables \\
  $\text{E}(\mathbb X)$   &   Expectation value of $\mathbb X$ \\
  $\text{Var}(\mathbb X)$   &   Variance of $\mathbb X$ \\
  $\text{Cov}(\mathbb X,\mathbb Y)$   &   Covariance of $\mathbb X$ and $\mathbb Y$ \\
\end{longtable}

\end{document}